\documentclass[doublespacing]{elsart}

\usepackage{amssymb}

\usepackage{theorem}
\theorembodyfont{\upshape}

\newtheorem{defi}{Definition}[section]

\newtheorem{propo}{Proposition}[section]
\newtheorem{lemma}[propo]{Lemma}

\newcommand{\beq}{\begin{equation}}
\newcommand{\eeq}{\end{equation}}
\newcommand{\beqs}{\begin{eqnarray}}
\newcommand{\eeqs}{\end{eqnarray}}

\begin{document}

\begin{frontmatter}

\title{Spanning forests on the Sierpinski gasket}

\author{Shu-Chiuan Chang}
\address{Department of Physics \\
National Cheng Kung University \\
Tainan 70101, Taiwan}
\ead{scchang@mail.ncku.edu.tw}

\author{Lung-Chi Chen}
\address{Department of Mathematics \\
Fu Jen Catholic University \\
Taipei 24205, Taiwan}
\ead{lcchen@math.fju.edu.tw}

\begin{abstract}

We present the numbers of spanning forests on the Sierpinski gasket $SG_d(n)$ at stage $n$ with dimension $d$ equal to two, three and four, and determine the asymptotic behaviors. The corresponding results on the generalized Sierpinski gasket $SG_{d,b}(n)$ with $d=2$ and $b=3,4$ are obtained. We also derive the upper bounds of the asymptotic growth constants for both $SG_d$ and $SG_{2,b}$.

\end{abstract}

\begin{keyword}
Spanning forests \sep Sierpinski gasket \sep recursion relations \sep exact solutions

\PACS 02.10.Ox
\end{keyword}
\end{frontmatter}

\section{Introduction}
\label{sectionI}

The enumeration of the number of spanning forests $N_{SF}(G)$ on a graph $G$ is a problem of interest in mathematics \cite{benjamini01,teranishi05} and physics \cite{sokal04}. It is well known that the number of spanning forests is given by the Tutte polynomial $T(G,x,y)$ evaluated at $x=2$, $y=1$ \cite{welsh}. Alternatively, it corresponds to a special $q \to 0$ limit of the partition function of the $q$-state Potts model in statistical mechanics \cite{sokal05}. Some recent studies on the enumeration of spanning forests and the calculation of their asymptotic growth constants on regular lattices were carried out in Refs. \cite{a,ta,hca,ka,s3a,Jacobsen05,deng06}. It is of interest to consider spanning forests on self-similar fractal lattices which have scaling invariance rather than translational invariance. Fractals are geometric structures of noninteger Hausdorff dimension realized by repeated construction of an elementary shape on progressively smaller length scales \cite{mandelbrot,Falconer}. A well-known example of fractal is the Sierpinski gasket. We shall derive the recursion relations for the numbers of spanning forests on the Sierpinski gasket with dimension equal to two, three and four, and determine the asymptotic growth constants. We shall also consider the number of spanning forests on the generalized Sierpinski gasket with dimension equal to two.

\section{Preliminaries}
\label{sectionII}

We first recall some relevant definitions for spanning forests and the Sierpinski gasket in this section. A connected graph (without loops) $G=(V,E)$ is defined by its vertex (site) and edge (bond) sets $V$ and $E$ \cite{bbook,fh}.  Let $v(G)=|V|$ be the number of vertices and $e(G)=|E|$ the number of edges in $G$.  A spanning subgraph $G^\prime$ is a subgraph of $G$ with the same vertex set $V$ and an edge set $E^\prime \subseteq E$. While a tree is a connected graph with no circuits, a spanning forest on $G$ is a spanning subgraph of $G$ that is a disjoint union of trees. That is, a subgraph of $G$ without any cycles, or an acyclic graph. Here an isolated vertex is considered as a tree. The degree or coordination number $k_i$ of a vertex $v_i \in V$ is the number of edges attached to it.  A $k$-regular graph is a graph with the property that each of its vertices has the same degree $k$. In general, one can associate an edge weight $x_{ij}$ to each edge connecting adjacent vertices $v_i$ and $v_j$ (see, for example \cite{alexander95}). For simplicity, all edge weights are set to one throughout this paper. 

When the number of spanning forests $N_{SF}(G)$ grows exponentially with $v(G)$ as $v(G) \to \infty$, there exists a constant $z_G$ describing this exponential growth \cite{burton93,lyons05}:
\beq
z_G = \lim_{v(G) \to \infty} \frac{\ln N_{SF}(G)}{v(G)}
\label{zdef}
\eeq
where $G$, when used as a subscript in this manner, implicitly refers to
the thermodynamic limit.

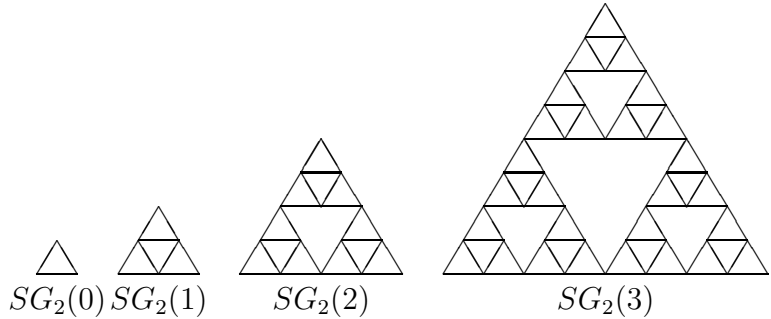
\begin{figure}[htbp]
\unitlength 0.9mm \hspace*{3mm}
\begin{picture}(108,40)
\put(0,0){\line(1,0){6}}
\put(0,0){\line(3,5){3}}
\put(6,0){\line(-3,5){3}}
\put(3,-4){\makebox(0,0){$SG_2(0)$}}
\put(12,0){\line(1,0){12}}
\put(12,0){\line(3,5){6}}
\put(24,0){\line(-3,5){6}}
\put(15,5){\line(1,0){6}}
\put(18,0){\line(3,5){3}}
\put(18,0){\line(-3,5){3}}
\put(18,-4){\makebox(0,0){$SG_2(1)$}}
\put(30,0){\line(1,0){24}}
\put(30,0){\line(3,5){12}}
\put(54,0){\line(-3,5){12}}
\put(36,10){\line(1,0){12}}
\put(42,0){\line(3,5){6}}
\put(42,0){\line(-3,5){6}}
\multiput(33,5)(12,0){2}{\line(1,0){6}}
\multiput(36,0)(12,0){2}{\line(3,5){3}}
\multiput(36,0)(12,0){2}{\line(-3,5){3}}
\put(39,15){\line(1,0){6}}
\put(42,10){\line(3,5){3}}
\put(42,10){\line(-3,5){3}}
\put(42,-4){\makebox(0,0){$SG_2(2)$}}
\put(60,0){\line(1,0){48}}
\put(72,20){\line(1,0){24}}
\put(60,0){\line(3,5){24}}
\put(84,0){\line(3,5){12}}
\put(84,0){\line(-3,5){12}}
\put(108,0){\line(-3,5){24}}
\put(66,10){\line(1,0){12}}
\put(90,10){\line(1,0){12}}
\put(78,30){\line(1,0){12}}
\put(72,0){\line(3,5){6}}
\put(96,0){\line(3,5){6}}
\put(84,20){\line(3,5){6}}
\put(72,0){\line(-3,5){6}}
\put(96,0){\line(-3,5){6}}
\put(84,20){\line(-3,5){6}}
\multiput(63,5)(12,0){4}{\line(1,0){6}}
\multiput(66,0)(12,0){4}{\line(3,5){3}}
\multiput(66,0)(12,0){4}{\line(-3,5){3}}
\multiput(69,15)(24,0){2}{\line(1,0){6}}
\multiput(72,10)(24,0){2}{\line(3,5){3}}
\multiput(72,10)(24,0){2}{\line(-3,5){3}}
\multiput(75,25)(12,0){2}{\line(1,0){6}}
\multiput(78,20)(12,0){2}{\line(3,5){3}}
\multiput(78,20)(12,0){2}{\line(-3,5){3}}
\put(81,35){\line(1,0){6}}
\put(84,30){\line(3,5){3}}
\put(84,30){\line(-3,5){3}}
\put(84,-4){\makebox(0,0){$SG_2(3)$}}
\end{picture}

\vspace*{5mm}
\caption{\footnotesize{The first four stages $n=0,1,2,3$ of the two-dimensional Sierpinski gasket $SG_2(n)$.}} 
\label{sgfig}
\end{figure}

The construction of the two-dimensional Sierpinski gasket $SG_2(n)$ at stage $n$ is shown in Fig. \ref{sgfig}. At stage $n=0$, it is an equilateral triangle; while stage $n+1$ is obtained by the juxtaposition of three $n$-stage structures. In general, the Sierpinski gaskets $SG_d$ can be built in any Euclidean dimension $d$ with fractal dimensionality $D=\ln(d+1)/\ln2$ \cite{Gefen81}. For the Sierpinski gasket $SG_d(n)$, the numbers of edges and vertices are given by 
\beq
e(SG_d(n)) = {d+1 \choose 2} (d+1)^n = \frac{d}{2} (d+1)^{n+1} \ ,
\label{e}
\eeq
\beq
v(SG_d(n)) = \frac{d+1}{2} [(d+1)^n+1] \ .
\label{v}
\eeq
Except the $(d+1)$ outmost vertices which have degree $d$, all other vertices of $SG_d(n)$ have degree $2d$. In the large $n$ limit, $SG_d$ is $2d$-regular. 

\bigskip

\bigskip

\begin{figure}[htbp]
\unitlength 0.9mm \hspace*{3mm}
\begin{picture}(108,45)
\put(0,0){\line(1,0){18}}
\put(3,5){\line(1,0){12}}
\put(6,10){\line(1,0){6}}
\put(0,0){\line(3,5){9}}
\put(6,0){\line(3,5){6}}
\put(12,0){\line(3,5){3}}
\put(18,0){\line(-3,5){9}}
\put(12,0){\line(-3,5){6}}
\put(6,0){\line(-3,5){3}}
\put(9,-4){\makebox(0,0){$SG_{2,3}(1)$}}
\put(24,0){\line(1,0){54}}
\put(33,15){\line(1,0){36}}
\put(42,30){\line(1,0){18}}
\put(24,0){\line(3,5){27}}
\put(42,0){\line(3,5){18}}
\put(60,0){\line(3,5){9}}
\put(78,0){\line(-3,5){27}}
\put(60,0){\line(-3,5){18}}
\put(42,0){\line(-3,5){9}}
\multiput(27,5)(18,0){3}{\line(1,0){12}}
\multiput(30,10)(18,0){3}{\line(1,0){6}}
\multiput(30,0)(18,0){3}{\line(3,5){6}}
\multiput(36,0)(18,0){3}{\line(3,5){3}}
\multiput(36,0)(18,0){3}{\line(-3,5){6}}
\multiput(30,0)(18,0){3}{\line(-3,5){3}}
\multiput(36,20)(18,0){2}{\line(1,0){12}}
\multiput(39,25)(18,0){2}{\line(1,0){6}}
\multiput(39,15)(18,0){2}{\line(3,5){6}}
\multiput(45,15)(18,0){2}{\line(3,5){3}}
\multiput(45,15)(18,0){2}{\line(-3,5){6}}
\multiput(39,15)(18,0){2}{\line(-3,5){3}}
\put(45,35){\line(1,0){12}}
\put(48,40){\line(1,0){6}}
\put(48,30){\line(3,5){6}}
\put(54,30){\line(3,5){3}}
\put(54,30){\line(-3,5){6}}
\put(48,30){\line(-3,5){3}}
\put(48,-4){\makebox(0,0){$SG_{2,3}(2)$}}
\put(84,0){\line(1,0){24}}
\put(87,5){\line(1,0){18}}
\put(90,10){\line(1,0){12}}
\put(93,15){\line(1,0){6}}
\put(84,0){\line(3,5){12}}
\put(90,0){\line(3,5){9}}
\put(96,0){\line(3,5){6}}
\put(102,0){\line(3,5){3}}
\put(108,0){\line(-3,5){12}}
\put(102,0){\line(-3,5){9}}
\put(96,0){\line(-3,5){6}}
\put(90,0){\line(-3,5){3}}
\put(96,-4){\makebox(0,0){$SG_{2,4}(1)$}}
\end{picture}

\vspace*{5mm}
\caption{\footnotesize{The generalized two-dimensional Sierpinski gasket $SG_{2,b}(n)$ with $b=3$ at stage $n=1, 2$ and $b=4$ at stage $n=1$.}} 
\label{sgbfig}
\end{figure}
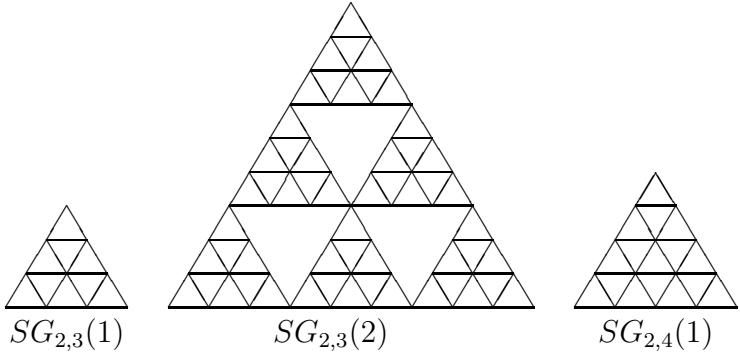

The Sierpinski gasket can be generalized, denoted as $SG_{d,b}(n)$, by introducing the side length $b$ which is an integer larger or equal to two \cite{Hilfer}. The generalized Sierpinski gasket at stage $n+1$ is constructed with $b$ layers of stage $n$ hypertetrahedrons. The two-dimensional $SG_{2,b}(n)$ with $b=3$ at stage $n=1, 2$ and $b=4$ at stage $n=1$ are illustrated in Fig. \ref{sgbfig}. The ordinary Sierpinski gasket $SG_d(n)$ corresponds to the $b=2$ case, where the index $b$ is neglected for simplicity. The Hausdorff dimension for $SG_{d,b}$ is given by $D=\ln {b+d-1 \choose d} / \ln b$ \cite{Hilfer}. Notice that $SG_{d,b}$ is not $k$-regular even in the thermodynamic limit.

\bigskip

\bigskip

\section{The number of spanning forests on $SG_2(n)$}
\label{sectionIII}

In this section we derive the asymptotic growth constant for the number of spanning forests on the two-dimensional Sierpinski gasket $SG_2(n)$ in detail. Let us start with the definitions of the quantities to be used.

\bigskip

\begin{defi} \label{defisg2} Consider the generalized two-dimensional Sierpinski gasket $SG_{2,b}(n)$ at stage $n$. (a) Define $f_{2,b}(n) \equiv N_{SF}(SG_{2,b}(n))$ as the number of spanning forests. (b) Define $t_{2,b}(n)$ as the number of spanning forests such that the three outmost vertices belong to one tree. (c) Define $ga_{2,b}(n)$, $gb_{2,b}(n)$, $gc_{2,b}(n)$ as the number of spanning forests such that one of the outmost vertices belongs to one tree and the other two outmost vertices belong to another tree. (d) Define $h_{2,b}(n)$ as the number of spanning forests such that each of the outmost vertices belongs to a different tree.
\end{defi}

\bigskip

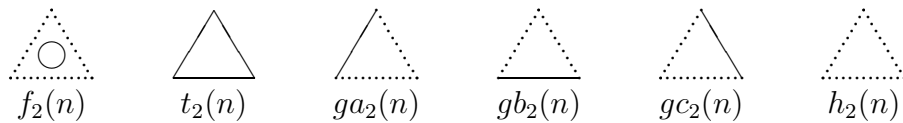
\begin{figure}[htbp]
\unitlength 1.8mm 
\begin{picture}(66,5)
\put(3,1.7){\circle{2}}
\multiput(0,0)(0.5,0){13}{\circle*{0.2}}
\multiput(0,0)(0.3,0.5){11}{\circle*{0.2}}
\multiput(6,0)(-0.3,0.5){11}{\circle*{0.2}}
\put(3,-2){\makebox(0,0){$f_2(n)$}}
\put(12,0){\line(1,0){6}}
\put(12,0){\line(3,5){3}}
\put(18,0){\line(-3,5){3}}
\put(15,-2){\makebox(0,0){$t_2(n)$}}
\multiput(24,0)(0.5,0){13}{\circle*{0.2}}
\put(24,0){\line(3,5){3}}
\multiput(30,0)(-0.3,0.5){11}{\circle*{0.2}}
\put(27,-2){\makebox(0,0){$ga_2(n)$}}
\put(36,0){\line(1,0){6}}
\multiput(36,0)(0.3,0.5){11}{\circle*{0.2}}
\multiput(42,0)(-0.3,0.5){11}{\circle*{0.2}}
\put(39,-2){\makebox(0,0){$gb_2(n)$}}
\multiput(48,0)(0.5,0){13}{\circle*{0.2}}
\multiput(48,0)(0.3,0.5){11}{\circle*{0.2}}
\put(54,0){\line(-3,5){3}}
\put(51,-2){\makebox(0,0){$gc_2(n)$}}
\multiput(60,0)(0.5,0){13}{\circle*{0.2}}
\multiput(60,0)(0.3,0.5){11}{\circle*{0.2}}
\multiput(66,0)(-0.3,0.5){11}{\circle*{0.2}}
\put(63,-2){\makebox(0,0){$h_2(n)$}}
\end{picture}

\vspace*{5mm}
\caption{\footnotesize{Illustration for the spanning subgraphs $f_2(n)$, $t_2(n)$, $ga_2(n)$, $gb_2(n)$, $gc_2(n)$ and $h_2(n)$. The two outmost vertices at the ends of a solid line belong to one tree, while the two outmost vertices at the ends of a dot line belong to separated trees.}} 
\label{ftghfig}
\end{figure}

Since we only consider ordinary Sierpinski gasket in this section, we use the notations $f_2(n)$, $t_2(n)$, $ga_2(n)$, $gb_2(n)$, $gc_2(n)$ and $h_2(n)$ for simplicity. They are illustrated in Fig. \ref{ftghfig}, where only the outmost vertices are shown. It is clear that the values $ga_2(n)$, $gb_2(n)$, $gc_2(n)$ are the same because of rotation symmetry, and we define $g_2(n) \equiv ga_2(n) = gb_2(n) = gc_2(n)$. Similarly for the generalized case, we define $g_{2,b}(n) \equiv ga_{2,b}(n) = gb_{2,b}(n) = gc_{2,b}(n)$. It follows that
\beq
f_2(n) = t_2(n)+3g_2(n)+h_2(n) \ .
\label{fsg2}
\eeq
The initial values at stage zero are $t_2(0)=3$, $g_2(0)=1$, $h_2(0)=1$ and $f_2(0)=7$. The purpose of this section is to obtain the asymptotic behavior of $f_2(n)$ as follows.
The three quantities $t_2(n)$, $g_2(n)$ and $h_2(n)$ satisfy recursion relations. 

\bigskip

\bigskip

\begin{lemma} \label{lemmasg2r} For any non-negative integer $n$,
\beq
f_2(n+1) = f_2^3(n) - [t_2(n)+g_2(n)]^3 \ , 
\label{feq}
\eeq
\beq
t_2(n+1) = 6t_2^2(n)g_2(n) + 3t_2(n)g_2^2(n) \ , 
\label{teq}
\eeq
\beqs
g_2(n+1) & = & t_2^2(n)h_2(n) + 2 t_2(n)g_2(n)h_2(n) + 7t_2(n)g_2^2(n) + 4g_2^3(n) \cr\cr
& & + g_2^2(n)h_2(n) \ , 
\label{geq}
\eeqs
\beqs
h_2(n+1) & = & 12t_2(n)g_2(n)h_2(n) + 14g_2^3(n) + 24g_2^2(n)h_2(n) + 9g_2(n)h_2^2(n) \cr\cr
& & + 3t_2(n)h_2^2(n) + h_2^3(n) \ .
\label{heq}
\eeqs
\end{lemma}

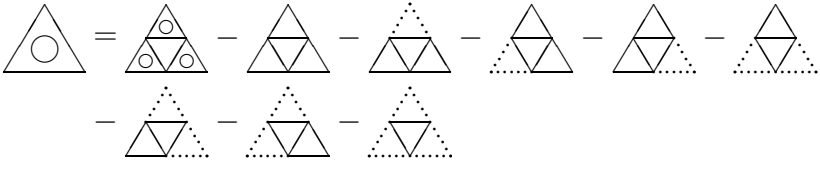
\begin{figure}[htbp]
\unitlength 0.9mm 
\begin{picture}(120,12)
\put(0,0){\line(1,0){12}}
\put(0,0){\line(3,5){6}}
\put(12,0){\line(-3,5){6}}
\put(6,3.4){\circle{4}}
\put(15,5){\makebox(0,0){$=$}}
\put(18,0){\line(1,0){12}}
\put(18,0){\line(3,5){6}}
\put(24,0){\line(3,5){3}}
\put(24,0){\line(-3,5){3}}
\put(21,5){\line(1,0){6}}
\put(30,0){\line(-3,5){6}}
\put(21,1.7){\circle{2}}
\put(27,1.7){\circle{2}}
\put(24,6.7){\circle{2}}
\put(33,5){\makebox(0,0){$-$}}
\put(36,0){\line(3,5){6}}
\put(36,0){\line(1,0){12}}
\put(42,0){\line(3,5){3}}
\put(42,0){\line(-3,5){3}}
\put(39,5){\line(1,0){6}}
\put(48,0){\line(-3,5){6}}
\put(51,5){\makebox(0,0){$-$}}
\put(57,5){\line(1,0){6}}
\put(54,0){\line(1,0){12}}
\multiput(54,0)(6,0){2}{\line(3,5){3}}
\multiput(60,0)(6,0){2}{\line(-3,5){3}}
\multiput(63,5)(-0.6,1){6}{\circle*{0.2}}
\multiput(57,5)(0.6,1){6}{\circle*{0.2}}
\put(69,5){\makebox(0,0){$-$}}
\put(78,0){\line(-3,5){3}}
\put(84,0){\line(-3,5){6}}
\multiput(78,0)(-3,5){2}{\line(1,0){6}}
\multiput(78,0)(-3,5){2}{\line(3,5){3}}
\multiput(72,0)(1,0){7}{\circle*{0.2}}
\multiput(72,0)(0.6,1){6}{\circle*{0.2}}
\put(87,5){\makebox(0,0){$-$}}
\put(90,0){\line(3,5){6}}
\put(96,0){\line(3,5){3}}
\multiput(90,0)(3,5){2}{\line(1,0){6}}
\multiput(96,0)(3,5){2}{\line(-3,5){3}}
\multiput(96,0)(1,0){7}{\circle*{0.2}}
\multiput(102,0)(-0.6,1){6}{\circle*{0.2}}
\put(105,5){\makebox(0,0){$-$}}
\put(111,5){\line(1,0){6}}
\multiput(114,0)(-3,5){2}{\line(3,5){3}}
\multiput(114,0)(3,5){2}{\line(-3,5){3}}
\multiput(108,0)(1,0){13}{\circle*{0.2}}
\multiput(108,0)(0.6,1){6}{\circle*{0.2}}
\multiput(120,0)(-0.6,1){6}{\circle*{0.2}}
\end{picture}

\unitlength 0.9mm 
\begin{picture}(66,12)
\put(15,5){\makebox(0,0){$-$}}
\put(24,0){\line(-3,5){3}}
\multiput(18,0)(3,5){2}{\line(1,0){6}}
\multiput(18,0)(6,0){2}{\line(3,5){3}}
\multiput(24,0)(1,0){7}{\circle*{0.2}}
\multiput(21,5)(0.6,1){6}{\circle*{0.2}}
\multiput(30,0)(-0.6,1){11}{\circle*{0.2}}
\put(33,5){\makebox(0,0){$-$}}
\put(42,0){\line(3,5){3}}
\multiput(42,0)(-3,5){2}{\line(1,0){6}}
\multiput(42,0)(6,0){2}{\line(-3,5){3}}
\multiput(36,0)(1,0){7}{\circle*{0.2}}
\multiput(36,0)(0.6,1){11}{\circle*{0.2}}
\multiput(45,5)(-0.6,1){6}{\circle*{0.2}}
\put(51,5){\makebox(0,0){$-$}}
\put(60,0){\line(3,5){3}}
\put(60,0){\line(-3,5){3}}
\put(57,5){\line(1,0){6}}
\multiput(54,0)(1,0){13}{\circle*{0.2}}
\multiput(54,0)(0.6,1){11}{\circle*{0.2}}
\multiput(66,0)(-0.6,1){11}{\circle*{0.2}}
\end{picture}

\caption{\footnotesize{Illustration for the expression of  $f_2(n+1)$.}} 
\label{ffig}
\end{figure}

{\sl Proof} \quad 
The Sierpinski gaskets $SG_2(n+1)$ is composed of three $SG_2(n)$ with three pairs of vertices identified. For the number $f_2(n+1)$, the unallowable configurations are those with a circuit, i.e., the two identified vertices of each $SG_2(n)$ belong to the same tree as illustrated in Fig. \ref{ffig}. Therefore, we have
\beq
f_2(n+1) = f_2^3(n) - [t_2(n)+ga_2(n)][t_2(n)+gb_2(n)][t_2(n)+gc_2(n)] \ .
\eeq
With the identity $ga_2(n)=gb_2(n)=gc_2(n)=g_2(n)$, Eq. (\ref{feq}) is verified.

\bigskip

\bigskip

\begin{figure}[htbp]
\unitlength 0.9mm 
\begin{picture}(73,12)
\put(0,0){\line(1,0){12}}
\put(0,0){\line(3,5){6}}
\put(12,0){\line(-3,5){6}}
\put(15,5){\makebox(0,0){$=$}}
\put(18,0){\line(1,0){6}}
\put(21,5){\line(1,0){6}}
\put(18,0){\line(3,5){6}}
\put(30,0){\line(-3,5){6}}
\put(24,0){\line(-3,5){3}}
\multiput(24,0)(1,0){7}{\circle*{0.2}}
\multiput(24,0)(0.6,1){6}{\circle*{0.2}}
\put(31,5){\makebox(0,0){$\times 3$}}
\put(36,5){\makebox(0,0){$+$}}
\put(39,0){\line(1,0){12}}
\put(42,5){\line(1,0){6}}
\put(39,0){\line(3,5){6}}
\multiput(45,0)(3,5){2}{\line(-3,5){3}}
\multiput(45,0)(0.6,1){6}{\circle*{0.2}}
\multiput(51,0)(-0.6,1){6}{\circle*{0.2}}
\put(52,5){\makebox(0,0){$\times 3$}}
\put(57,5){\makebox(0,0){$+$}}
\put(60,0){\line(3,5){6}}
\put(72,0){\line(-3,5){6}}
\put(63,5){\line(1,0){6}}
\multiput(66,0)(-0.6,1){6}{\circle*{0.2}}
\multiput(66,0)(0.6,1){6}{\circle*{0.2}}
\multiput(60,0)(1,0){13}{\circle*{0.2}}
\put(73,5){\makebox(0,0){$\times 3$}}
\end{picture}

\caption{\footnotesize{Illustration for the expression of  $t_2(n+1)$. The multiplication of three on the right-hand-side corresponds to the three possible orientations of $SG_2(n+1)$.}} 
\label{tfig}
\end{figure}
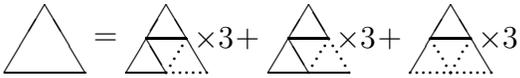

The number $t_2(n+1)$ consists of six configurations where two of the $SG_2(n)$ are in the $t_2(n)$ status and the other one is in the $g_2(n)$ status, and three configurations where one of the $SG_2(n)$ is in the $t_2(n)$ status and the other two are in the $g_2(n)$ status as illustrated in Fig. \ref{tfig}. Therefore, we have
\beqs
t_2(n+1) & = & 2t_2^2(n)[ga_2(n)+gb_2(n)+gc_2(n)] + t_2(n)ga_2(n)gb_2(n) \cr\cr
& & + t_2(n)ga_2(n)gc_2(n) + t_2(n)gb_2(n)gc_2(n) \ .
\eeqs
With the identity $ga_2(n)=gb_2(n)=gc_2(n)=g_2(n)$, Eq. (\ref{teq}) is verified.

\bigskip

\bigskip

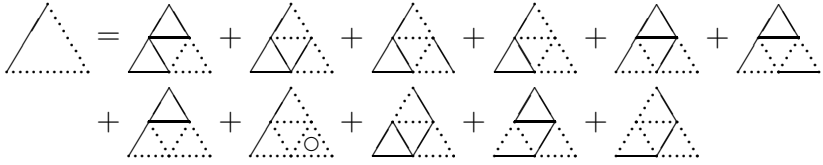
\begin{figure}[htbp]
\unitlength 0.9mm 
\begin{picture}(120,12)
\put(0,0){\line(3,5){6}}
\multiput(0,0)(1,0){13}{\circle*{0.2}}
\multiput(12,0)(-0.6,1){11}{\circle*{0.2}}
\put(15,5){\makebox(0,0){$=$}}
\put(18,0){\line(3,5){6}}
\multiput(18,0)(3,5){2}{\line(1,0){6}}
\multiput(24,0)(3,5){2}{\line(-3,5){3}}
\multiput(24,0)(1,0){7}{\circle*{0.2}}
\multiput(24,0)(0.6,1){6}{\circle*{0.2}}
\multiput(30,0)(-0.6,1){6}{\circle*{0.2}}
\put(33,5){\makebox(0,0){$+$}}
\put(36,0){\line(3,5){6}}
\put(36,0){\line(1,0){6}}
\put(42,0){\line(3,5){3}}
\put(42,0){\line(-3,5){3}}
\multiput(42,0)(1,0){7}{\circle*{0.2}}
\multiput(39,5)(1,0){7}{\circle*{0.2}}
\multiput(48,0)(-0.6,1){11}{\circle*{0.2}}
\put(51,5){\makebox(0,0){$+$}}
\put(54,0){\line(3,5){6}}
\put(54,0){\line(1,0){6}}
\multiput(60,0)(6,0){2}{\line(-3,5){3}}
\multiput(60,0)(1,0){7}{\circle*{0.2}}
\multiput(60,0)(0.6,1){6}{\circle*{0.2}}
\multiput(57,5)(1,0){7}{\circle*{0.2}}
\multiput(63,5)(-0.6,1){6}{\circle*{0.2}}
\put(69,5){\makebox(0,0){$+$}}
\put(72,0){\line(3,5){6}}
\put(72,0){\line(1,0){6}}
\put(78,0){\line(-3,5){3}}
\multiput(78,0)(1,0){7}{\circle*{0.2}}
\multiput(75,5)(1,0){7}{\circle*{0.2}}
\multiput(78,0)(0.6,1){6}{\circle*{0.2}}
\multiput(84,0)(-0.6,1){11}{\circle*{0.2}}
\put(87,5){\makebox(0,0){$+$}}
\put(90,0){\line(3,5){6}}
\put(93,5){\line(1,0){6}}
\put(96,0){\line(3,5){3}}
\put(99,5){\line(-3,5){3}}
\multiput(90,0)(1,0){13}{\circle*{0.2}}
\multiput(96,0)(-0.6,1){6}{\circle*{0.2}}
\multiput(102,0)(-0.6,1){6}{\circle*{0.2}}
\put(105,5){\makebox(0,0){$+$}}
\put(108,0){\line(3,5){6}}
\put(117,5){\line(-3,5){3}}
\multiput(114,0)(-3,5){2}{\line(1,0){6}}
\multiput(108,0)(1,0){7}{\circle*{0.2}}
\multiput(114,0)(-0.6,1){6}{\circle*{0.2}}
\multiput(114,0)(0.6,1){6}{\circle*{0.2}}
\multiput(120,0)(-0.6,1){6}{\circle*{0.2}}
\end{picture}

\unitlength 0.9mm 
\begin{picture}(102,12)
\put(15,5){\makebox(0,0){$+$}}
\put(18,0){\line(3,5){6}}
\put(27,5){\line(-3,5){3}}
\put(21,5){\line(1,0){6}}
\multiput(18,0)(1,0){13}{\circle*{0.2}}
\multiput(24,0)(-0.6,1){6}{\circle*{0.2}}
\multiput(24,0)(0.6,1){6}{\circle*{0.2}}
\multiput(30,0)(-0.6,1){6}{\circle*{0.2}}
\put(33,5){\makebox(0,0){$+$}}
\put(36,0){\line(3,5){6}}
\multiput(36,0)(1,0){13}{\circle*{0.2}}
\multiput(39,5)(1,0){7}{\circle*{0.2}}
\multiput(42,0)(0.6,1){6}{\circle*{0.2}}
\multiput(42,0)(-0.6,1){6}{\circle*{0.2}}
\multiput(48,0)(-0.6,1){11}{\circle*{0.2}}
\put(45,1.7){\circle{2}}
\put(51,5){\makebox(0,0){$+$}}
\put(54,0){\line(3,5){3}}
\put(54,0){\line(1,0){6}}
\put(60,0){\line(3,5){3}}
\multiput(60,0)(3,5){2}{\line(-3,5){3}}
\multiput(60,0)(1,0){7}{\circle*{0.2}}
\multiput(57,5)(1,0){7}{\circle*{0.2}}
\multiput(57,5)(0.6,1){6}{\circle*{0.2}}
\multiput(66,0)(-0.6,1){6}{\circle*{0.2}}
\put(69,5){\makebox(0,0){$+$}}
\multiput(72,0)(3,5){2}{\line(1,0){6}}
\multiput(78,0)(-3,5){2}{\line(3,5){3}}
\put(81,5){\line(-3,5){3}}
\multiput(78,0)(1,0){7}{\circle*{0.2}}
\multiput(72,0)(0.6,1){6}{\circle*{0.2}}
\multiput(78,0)(-0.6,1){6}{\circle*{0.2}}
\multiput(84,0)(-0.6,1){6}{\circle*{0.2}}
\put(87,5){\makebox(0,0){$+$}}
\put(90,0){\line(1,0){6}}
\put(96,0){\line(3,5){3}}
\put(99,5){\line(-3,5){3}}
\multiput(90,0)(0.6,1){11}{\circle*{0.2}}
\multiput(96,0)(1,0){7}{\circle*{0.2}}
\multiput(93,5)(1,0){7}{\circle*{0.2}}
\multiput(96,0)(-0.6,1){6}{\circle*{0.2}}
\multiput(102,0)(-0.6,1){6}{\circle*{0.2}}
\end{picture}

\caption{\footnotesize{Illustration for the expression of  $ga_2(n+1)$.}} 
\label{gfig}
\end{figure}

Similarly, $ga_2(n+1)$ for $SG_2(n+1)$ can be obtained with appropriated configurations of its three constituting $SG_2(n)$ as illustrated in Fig. \ref{gfig}. Thus,
\beqs
ga_2(n+1) & = & t_2^2(n)h_2(n) + t_2(n)ga_2(n)[ga_2(n)+gc_2(n)+h_2(n)] \cr\cr
& & + t_2(n)ga_2(n)[ga_2(n)+gb_2(n)+h_2(n)] + f_2(n)ga_2^2(n) \cr\cr
& & + t_2(n)ga_2(n)gc_2(n) + t_2(n)ga_2(n)gb_2(n) \cr\cr
& & + ga_2(n)gb_2(n)gc_2(n) \ .
\eeqs
With the identity $ga_2(n)=gb_2(n)=gc_2(n)=g_2(n)$ and Eq. (\ref{fsg2}), Eq. (\ref{geq}) is verified.

\bigskip

\bigskip

\begin{figure}[htbp]
\unitlength 0.9mm 
\begin{picture}(136,12)
\multiput(0,0)(1,0){13}{\circle*{0.2}}
\multiput(0,0)(0.6,1){11}{\circle*{0.2}}
\multiput(12,0)(-0.6,1){11}{\circle*{0.2}}
\put(15,5){\makebox(0,0){$=$}}
\put(18,0){\line(1,0){6}}
\put(18,0){\line(3,5){3}}
\multiput(24,0)(6,0){2}{\line(-3,5){3}}
\multiput(24,0)(1,0){7}{\circle*{0.2}}
\multiput(21,5)(1,0){7}{\circle*{0.2}}
\multiput(24,0)(0.6,1){6}{\circle*{0.2}}
\multiput(21,5)(0.6,1){6}{\circle*{0.2}}
\multiput(27,5)(-0.6,1){6}{\circle*{0.2}}
\put(31,5){\makebox(0,0){$\times 3$}}
\put(36,5){\makebox(0,0){$+$}}
\put(39,0){\line(1,0){6}}
\put(45,0){\line(-3,5){3}}
\multiput(39,0)(6,0){2}{\line(3,5){3}}
\multiput(45,0)(1,0){7}{\circle*{0.2}}
\multiput(42,5)(1,0){7}{\circle*{0.2}}
\multiput(42,5)(0.6,1){6}{\circle*{0.2}}
\multiput(51,0)(-0.6,1){11}{\circle*{0.2}}
\put(52,5){\makebox(0,0){$\times 3$}}
\put(57,5){\makebox(0,0){$+$}}
\put(60,0){\line(1,0){6}}
\put(60,0){\line(3,5){3}}
\multiput(66,0)(3,5){2}{\line(-3,5){3}}
\multiput(66,0)(1,0){7}{\circle*{0.2}}
\multiput(63,5)(1,0){7}{\circle*{0.2}}
\multiput(66,0)(0.6,1){6}{\circle*{0.2}}
\multiput(63,5)(0.6,1){6}{\circle*{0.2}}
\multiput(72,0)(-0.6,1){6}{\circle*{0.2}}
\put(73,5){\makebox(0,0){$\times 3$}}
\put(78,5){\makebox(0,0){$+$}}
\put(81,0){\line(3,5){3}}
\put(87,0){\line(-3,5){3}}
\multiput(81,0)(3,5){2}{\line(1,0){6}}
\multiput(87,0)(1,0){7}{\circle*{0.2}}
\multiput(87,0)(0.6,1){6}{\circle*{0.2}}
\multiput(84,5)(0.6,1){6}{\circle*{0.2}}
\multiput(93,0)(-0.6,1){11}{\circle*{0.2}}
\put(94,5){\makebox(0,0){$\times 3$}}
\put(99,5){\makebox(0,0){$+$}}
\put(108,0){\line(3,5){3}}
\multiput(108,0)(3,5){2}{\line(-3,5){3}}
\multiput(102,0)(1,0){13}{\circle*{0.2}}
\multiput(102,0)(0.6,1){11}{\circle*{0.2}}
\multiput(105,5)(1,0){7}{\circle*{0.2}}
\multiput(114,0)(-0.6,1){6}{\circle*{0.2}}
\put(115,5){\makebox(0,0){$\times 3$}}
\put(120,5){\makebox(0,0){$+$}}
\put(129,0){\line(-3,5){3}}
\multiput(129,0)(-3,5){2}{\line(3,5){3}}
\multiput(123,0)(1,0){13}{\circle*{0.2}}
\multiput(135,0)(-0.6,1){11}{\circle*{0.2}}
\multiput(126,5)(1,0){7}{\circle*{0.2}}
\multiput(123,0)(0.6,1){6}{\circle*{0.2}}
\put(136,5){\makebox(0,0){$\times 3$}}
\end{picture}

\unitlength 0.9mm 
\begin{picture}(135,12)
\put(15,5){\makebox(0,0){$+$}}
\put(21,5){\line(3,5){3}}
\multiput(24,0)(6,0){2}{\line(-3,5){3}}
\multiput(18,0)(1,0){13}{\circle*{0.2}}
\multiput(21,5)(1,0){7}{\circle*{0.2}}
\multiput(18,0)(0.6,1){6}{\circle*{0.2}}
\multiput(24,0)(0.6,1){6}{\circle*{0.2}}
\multiput(27,5)(-0.6,1){6}{\circle*{0.2}}
\put(31,5){\makebox(0,0){$\times 3$}}
\put(36,5){\makebox(0,0){$+$}}
\put(48,5){\line(-3,5){3}}
\multiput(39,0)(6,0){2}{\line(3,5){3}}
\multiput(39,0)(1,0){13}{\circle*{0.2}}
\multiput(42,5)(1,0){7}{\circle*{0.2}}
\multiput(42,5)(0.6,1){6}{\circle*{0.2}}
\multiput(45,0)(-0.6,1){6}{\circle*{0.2}}
\multiput(51,0)(-0.6,1){6}{\circle*{0.2}}
\put(52,5){\makebox(0,0){$\times 3$}}
\put(57,5){\makebox(0,0){$+$}}
\put(60,0){\line(1,0){6}}
\put(63,5){\line(3,5){3}}
\put(72,0){\line(-3,5){3}}
\multiput(60,0)(0.6,1){6}{\circle*{0.2}}
\multiput(66,0)(0.6,1){6}{\circle*{0.2}}
\multiput(66,0)(1,0){7}{\circle*{0.2}}
\multiput(63,5)(1,0){7}{\circle*{0.2}}
\multiput(66,0)(-0.6,1){6}{\circle*{0.2}}
\multiput(69,5)(-0.6,1){6}{\circle*{0.2}}
\put(78,5){\makebox(0,0){$+$}}
\put(87,0){\line(1,0){6}}
\put(81,0){\line(3,5){3}}
\put(90,5){\line(-3,5){3}}
\multiput(81,0)(1,0){7}{\circle*{0.2}}
\multiput(84,5)(1,0){7}{\circle*{0.2}}
\multiput(84,5)(0.6,1){6}{\circle*{0.2}}
\multiput(87,0)(0.6,1){6}{\circle*{0.2}}
\multiput(87,0)(-0.6,1){6}{\circle*{0.2}}
\multiput(93,0)(-0.6,1){6}{\circle*{0.2}}
\put(99,5){\makebox(0,0){$+$}}
\put(102,0){\line(1,0){6}}
\put(102,0){\line(3,5){3}}
\put(108,0){\line(-3,5){3}}
\multiput(108,0)(1,0){7}{\circle*{0.2}}
\multiput(105,5)(1,0){7}{\circle*{0.2}}
\multiput(108,0)(0.6,1){6}{\circle*{0.2}}
\multiput(105,5)(0.6,1){6}{\circle*{0.2}}
\multiput(114,0)(-0.6,1){11}{\circle*{0.2}}
\put(115,5){\makebox(0,0){$\times 3$}}
\put(120,5){\makebox(0,0){$+$}}
\multiput(123,0)(1,0){13}{\circle*{0.2}}
\multiput(123,0)(0.6,1){11}{\circle*{0.2}}
\multiput(135,0)(-0.6,1){11}{\circle*{0.2}}
\multiput(129,0)(0.6,1){6}{\circle*{0.2}}
\multiput(129,0)(-0.6,1){6}{\circle*{0.2}}
\multiput(126,5)(1,0){7}{\circle*{0.2}}
\end{picture}

\unitlength 0.9mm 
\begin{picture}(136,12)
\put(15,5){\makebox(0,0){$+$}}
\put(18,0){\line(3,5){3}}
\multiput(18,0)(1,0){13}{\circle*{0.2}}
\multiput(24,0)(0.6,1){6}{\circle*{0.2}}
\multiput(21,5)(0.6,1){6}{\circle*{0.2}}
\multiput(21,5)(1,0){7}{\circle*{0.2}}
\multiput(24,0)(-0.6,1){6}{\circle*{0.2}}
\multiput(30,0)(-0.6,1){11}{\circle*{0.2}}
\put(31,5){\makebox(0,0){$\times 3$}}
\put(36,5){\makebox(0,0){$+$}}
\put(39,0){\line(1,0){6}}
\multiput(39,0)(0.6,1){11}{\circle*{0.2}}
\multiput(42,5)(1,0){7}{\circle*{0.2}}
\multiput(45,0)(0.6,1){6}{\circle*{0.2}}
\multiput(45,0)(1,0){7}{\circle*{0.2}}
\multiput(45,0)(-0.6,1){6}{\circle*{0.2}}
\multiput(51,0)(-0.6,1){11}{\circle*{0.2}}
\put(52,5){\makebox(0,0){$\times 3$}}
\put(57,5){\makebox(0,0){$+$}}
\put(66,0){\line(-3,5){3}}
\multiput(60,0)(1,0){13}{\circle*{0.2}}
\multiput(60,0)(0.6,1){11}{\circle*{0.2}}
\multiput(72,0)(-0.6,1){11}{\circle*{0.2}}
\multiput(63,5)(1,0){7}{\circle*{0.2}}
\multiput(66,0)(0.6,1){6}{\circle*{0.2}}
\put(73,5){\makebox(0,0){$\times 3$}}
\put(78,5){\makebox(0,0){$+$}}
\put(87,0){\line(3,5){3}}
\put(84,5){\line(3,5){3}}
\multiput(81,0)(1,0){13}{\circle*{0.2}}
\multiput(84,5)(1,0){7}{\circle*{0.2}}
\multiput(81,0)(0.6,1){6}{\circle*{0.2}}
\multiput(87,0)(-0.6,1){6}{\circle*{0.2}}
\multiput(93,0)(-0.6,1){11}{\circle*{0.2}}
\put(94,5){\makebox(0,0){$\times 3$}}
\put(99,5){\makebox(0,0){$+$}}
\put(108,0){\line(1,0){6}}
\put(105,5){\line(3,5){3}}
\multiput(102,0)(1,0){7}{\circle*{0.2}}
\multiput(105,5)(1,0){7}{\circle*{0.2}}
\multiput(102,0)(0.6,1){6}{\circle*{0.2}}
\multiput(108,0)(0.6,1){6}{\circle*{0.2}}
\multiput(108,0)(-0.6,1){6}{\circle*{0.2}}
\multiput(114,0)(-0.6,1){11}{\circle*{0.2}}
\put(115,5){\makebox(0,0){$\times 3$}}
\put(120,5){\makebox(0,0){$+$}}
\put(135,0){\line(-3,5){3}}
\put(126,5){\line(3,5){3}}
\multiput(123,0)(1,0){13}{\circle*{0.2}}
\multiput(123,0)(0.6,1){6}{\circle*{0.2}}
\multiput(132,5)(-0.6,1){6}{\circle*{0.2}}
\multiput(129,0)(0.6,1){6}{\circle*{0.2}}
\multiput(129,0)(-0.6,1){6}{\circle*{0.2}}
\multiput(126,5)(1,0){7}{\circle*{0.2}}
\put(136,5){\makebox(0,0){$\times 3$}}
\end{picture}

\unitlength 0.9mm 
\begin{picture}(115,12)
\put(15,5){\makebox(0,0){$+$}}
\put(21,5){\line(1,0){6}}
\put(24,0){\line(3,5){3}}
\multiput(18,0)(1,0){13}{\circle*{0.2}}
\multiput(18,0)(0.6,1){11}{\circle*{0.2}}
\multiput(30,0)(-0.6,1){11}{\circle*{0.2}}
\multiput(24,0)(-0.6,1){6}{\circle*{0.2}}
\put(31,5){\makebox(0,0){$\times 3$}}
\put(36,5){\makebox(0,0){$+$}}
\put(45,0){\line(1,0){6}}
\put(42,5){\line(1,0){6}}
\multiput(39,0)(0.6,1){11}{\circle*{0.2}}
\multiput(39,0)(1,0){7}{\circle*{0.2}}
\multiput(45,0)(0.6,1){6}{\circle*{0.2}}
\multiput(45,0)(-0.6,1){6}{\circle*{0.2}}
\multiput(51,0)(-0.6,1){11}{\circle*{0.2}}
\put(52,5){\makebox(0,0){$\times 3$}}
\put(57,5){\makebox(0,0){$+$}}
\put(63,5){\line(1,0){6}}
\put(72,0){\line(-3,5){3}}
\multiput(60,0)(1,0){13}{\circle*{0.2}}
\multiput(60,0)(0.6,1){11}{\circle*{0.2}}
\multiput(66,0)(0.6,1){6}{\circle*{0.2}}
\multiput(66,0)(-0.6,1){6}{\circle*{0.2}}
\multiput(69,5)(-0.6,1){6}{\circle*{0.2}}
\put(73,5){\makebox(0,0){$\times 3$}}
\put(78,5){\makebox(0,0){$+$}}
\put(87,0){\line(3,5){3}}
\put(90,5){\line(-3,5){3}}
\multiput(81,0)(1,0){13}{\circle*{0.2}}
\multiput(81,0)(0.6,1){11}{\circle*{0.2}}
\multiput(84,5)(1,0){7}{\circle*{0.2}}
\multiput(87,0)(-0.6,1){6}{\circle*{0.2}}
\multiput(93,0)(-0.6,1){6}{\circle*{0.2}}
\put(94,5){\makebox(0,0){$\times 3$}}
\put(99,5){\makebox(0,0){$+$}}
\put(108,0){\line(1,0){6}}
\put(111,5){\line(-3,5){3}}
\multiput(102,0)(1,0){7}{\circle*{0.2}}
\multiput(102,0)(0.6,1){11}{\circle*{0.2}}
\multiput(105,5)(1,0){7}{\circle*{0.2}}
\multiput(108,0)(0.6,1){6}{\circle*{0.2}}
\multiput(108,0)(-0.6,1){6}{\circle*{0.2}}
\multiput(114,0)(-0.6,1){6}{\circle*{0.2}}
\put(115,5){\makebox(0,0){$\times 3$}}
\end{picture}

\caption{\footnotesize{Illustration for the expression of  $h_2(n+1)$. The multiplication of three on the right-hand-side corresponds to the three possible orientations $SG_2(n+1)$.}} 
\label{hfig}
\end{figure}
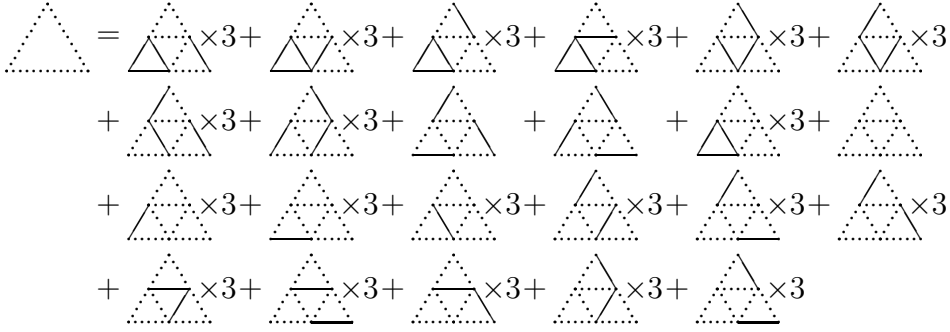

Finally, $h_2(n+1)$ is the summation of appropriated configurations as illustrated in Fig. \ref{hfig}, so that 
\beqs
& & h_2(n+1) \cr\cr
& = & 4t_2(n)h_2(n)[ga_2(n)+gb_2(n)+gc_2(n)] \cr\cr
& & + 2gc_2(n)ga_2(n)[gc_2(n)+ga_2(n)] + 2ga_2(n)gb_2(n)[ga_2(n)+gb_2(n)] \cr\cr
& & + 2gb_2(n)gc_2(n)[gb_2(n)+gc_2(n)] + 2ga_2(n)gb_2(n)gc_2(n) \cr\cr
& & + 3t_2(n)h_2^2(n) + h_2^3(n) + 3[ga_2(n)+gb_2(n)+gc_2(n)]h_2^2(n) \cr\cr
& & + \{3[ga_2(n)+gb_2(n)+gc_2(n)]^2-ga_2^2(n)-gb_2^2(n)-gc_2^2(n)\}h_2(n) \ .
\eeqs
With the identity $ga_2(n)=gb_2(n)=gc_2(n)=g_2(n)$, Eq. (\ref{heq}) is verified.
 
\bigskip

\bigskip

Eq. (\ref{feq}) can also be obtained by substituting Eqs. (\ref{teq}), (\ref{geq}) and (\ref{heq}) into Eq. (\ref{fsg2}). \ $\Box$

\bigskip

The values of $f_2(n)$, $t_2(n)$, $g_2(n)$, $h_2(n)$ for small $n$ can be evaluated recursively by Eqs. (\ref{feq}), (\ref{teq}), (\ref{geq}), (\ref{heq}) as listed in Table \ref{tablesg2}. These numbers grow exponentially, and do not have simple integer factorizations, in contrast to the corresponding results for the number of spanning trees \cite{sts}. To estimate the value of the asymptotic growth constant defined in Eq. (\ref{zdef}), we need the following lemma.

\bigskip

\begin{table}[htbp]
\caption{\label{tablesg2} The first few values of $f_2(n)$, $t_2(n)$, $g_2(n)$, $h_2(n)$.}
\begin{center}
\begin{tabular}{|c||r|r|r|r|}
\hline\hline 
$n$      & 0 &   1 &          2 &                             3 \\ \hline\hline 
$f_2(n)$ & 7 & 279 & 20,592,775 & 8,696,126,758,781,951,722,199 \\ \hline 
$t_2(n)$ & 3 &  63 &  1,294,083 &    36,212,372,367,917,382,063 \\ \hline 
$g_2(n)$ & 1 &  41 &  2,022,893 &   215,741,040,104,979,715,185 \\ \hline 
$h_2(n)$ & 1 &  93 & 13,230,013 & 8,012,691,266,099,095,194,581 \\ \hline\hline 
\end{tabular}
\end{center}
\end{table}

\bigskip

\begin{lemma} \label{lemmasg2b} The asymptotic growth constant for the number of spanning forests on $SG_2(n)$ is bounded:
\beq
\frac{2}{3^{m+1}} \ln h_2(m) < z_{SG_2} < \frac{2}{3^{m+1}} \ln f_2(m) \ ,
\label{zsg2}
\eeq
where $m$ is a positive integer.
\end{lemma}

{\sl Proof} \quad 
We first show that the ratio $t_2(n)/g_2(n)$ is a strictly decreasing sequence. By Eqs. (\ref{teq}) and (\ref{geq}), we have
\beqs
\lefteqn{\frac{t_2(n+1)}{g_2(n+1)}} \cr\cr
& = & \frac {6t_2^2(n)g_2(n)+3t_2(n)g_2^2(n)} {t_2^2(n)h_2(n) + 2t_2(n)g_2(n)h_2(n) + 7t_2(n)g_2^2(n) + 4g_2^3(n)+g_2^2(n)h_2(n)} \cr\cr
& < & \frac {t_2(n) [6t_2(n)g_2(n)+3g_2^2(n)]} {g_2(n)[ 7t_2(n)g_2(n)+4g_2^2(n)]} < \frac{6t_2(n)}{7g_2(n)} \ .
\eeqs
From the values in Table \ref{tablesg2}, $t_2(n)/g_2(n)$ is less than one for $n>1$. It is clear that this ratio approaches to zero as $n$ increases. Similarly, $g_2(n)/h_2(n)$ is also a strictly decreasing sequence by Eqs. (\ref{geq}) and (\ref{heq}).
\beqs
\lefteqn{\frac{g_2(n+1)}{h_2(n+1)}} \cr\cr & < & \frac {3t_2(n)g_2(n)h_2(n) + 7t_2(n)g_2^2(n) + 4g_2^3(n)+g_2^2(n)h_2(n)} {12t_2(n)g_2(n)h_2(n) + 14g_2^3(n) + 24g_2^2(n)h_2(n) + 9g_2(n)h_2^2(n) + 4t_2(n)h_2^2(n)} \cr\cr
& < & \frac {g_2(n)[3t_2(n)h_2(n)+7t_2(n)g_2(n)+4g_2^2(n)+g_2(n)h_2(n)]} {h_2(n)[4t_2(n)h_2(n)+12t_2(n)g_2(n)+24g_2^2(n)+9g_2(n)h_2(n)]} \cr\cr
& < & \frac{3g_2(n)}{4h_2(n)} \qquad \rm{for} \ n>1 \ ,
\eeqs
where we have used the fact that $t_2(n) < g_2(n) < h_2(n)$ for $n>1$.
Again, $g_2(n)/h_2(n)$ approaches to zero as $n$ increases. The relation $t_2(n) \ll g_2(n) \ll h_2(n)$ for large $n$ is expectable since it is rare to keep the three outmost vertices of $SG_2(n)$ in the same tree for $t_2(n)$ and $h_2(n)$ should dominate when $n$ becomes large. In fact, both $f_2(n)$ and $g_2(n)$ are negligible compared with $h_2(n)$ such that $f_2(n) \sim h_2(n)$ for large $n$. By Eqs. (\ref{feq}) and (\ref{heq}), we have the upper and lower bounds for $f_2(n)$:
\beq
h_2^3(n-1) < h_2(n) < f_2(n) < f_2^3(n-1) \ ,
\eeq
such that
\beq
h_2(m)^{3^{n-m}} < f_2(n) < f_2(m)^{3^{n-m}} \ ,
\eeq
where $m$ is a fixed integer. With the definition for $z_{SG_2}$ given in Eq. (\ref{zdef}) and the number of vertices of $SG_2(n)$ is $3(3^n+1)/2$ by Eq. (\ref{v}), the proof is completed.
\ $\Box$

\bigskip

\begin{propo} \label{proposg2} The asymptotic growth constant for the number of spanning forests on the two-dimensional Sierpinski gasket $SG_2(n)$ in the large $n$ limit is $z_{SG_2}=1.24733719931...$.

\end{propo}

{\sl Proof} \quad 
Define ratios $\alpha (n) \equiv t_2(n)/f_2(n)$ and $\beta (n) \equiv g_2(n)/f_2(n)$. By Eq. (\ref{fsg2}), it is clear that $0 \le \alpha (n) + \beta (n) < 1$. According to Lemma \ref{lemmasg2b}, $\alpha (n) + \beta (n)$ is a strictly decreasing sequence. By Eq. (\ref{feq}), let us define $r(n) \equiv f_2(n)/f_2^3(n-1) = 1-[\alpha (n-1) + \beta (n-1)]^3$ for positive integer $n$ . It follows that
\beqs
\ln f_2(n) & = & 3\ln f_2(n-1) + \ln r(n) = ... \cr\cr
& = & 3^{n-m} \ln f_2(m) + \sum_{j=m+1}^n 3^{n-j} \ln r(j) \cr\cr
& > & 3^{n-m} \ln f_2(m) + \left ( \frac{3^{n-m}-1}{2} \right ) \ln r(m+1) \ .
\eeqs
Divide this equation by $3(3^n+1)/2$ and take the limit $n \to \infty$, the difference between the upper bound in Eq. (\ref{zsg2}) and the asymptotic growth constant is bounded:
\beq
\frac{2}{3^{m+1}} \ln f_2(m) - z_{SG_2} \le \frac{-1}{3^{m+1}} \ln \left ( 1-[\alpha (m) + \beta (m)]^3 \right ) \ .
\label{zsg2d}
\eeq
When $m$ is as small as three, the right-hand-side of Eq. (\ref{zsg2d}) is about $3\times 10^{-7}$ by the values given in Table \ref{tablesg2}. Similarly, it can be shown that the difference between $z_{SG_2}$ and the lower bound (left-hand-side of Eq. (\ref{zsg2})) quickly converges to zero as $m$ increases. In another word, the numerical values of $\ln f_2(m)$ and $\ln h_2(m)$ are almost the same except the first few $m$, and the upper and lower bounds in Eq. (\ref{zsg2}) converge to the quoted value of $z_{SG_2}$. In fact, one obtains the numerical value of $z_{SG_2}$ with more than a hundred significant figures accurate when $m$ is equal to eight.
\ $\Box$

\bigskip

\section{The number of spanning forests on $SG_{2,b}(n)$ with $b=3,4$} 
\label{sectionIV}

The method given in the previous section can be applied to the number of spanning forests on $SG_{d,b}(n)$ with larger values of $d$ and $b$. The number of configurations to be considered increases as $d$ and $b$ increase, and the recursion relations must be derived individually for each $d$ and $b$. 
In this section, we consider the generalized two-dimensional Sierpinski gasket $SG_{2,b}(n)$ with the number of layers $b$ equal to three and four. 
For $SG_{2,3}(n)$, the numbers of edges and vertices are given by 
\beq
e(SG_{2,3}(n)) = 3 \times 6^n \ ,
\label{esg23}
\eeq
\beq
v(SG_{2,3}(n)) = \frac{7 \times 6^n + 8}{5} \ ,
\label{vsg23}
\eeq
where the three outmost vertices have degree two. There are $(6^n-1)/5$ vertices of $SG_{2,3}(n)$ with degree six and $6(6^n-1)/5$ vertices with degree four. By Definition \ref{defisg2}, the number of spanning forests is $f_{2,3}(n) = t_{2,3}(n)+3g_{2,3}(n)+h_{2,3}(n)$. The initial values are the same as for $SG_2$: $t_{2,3}(0)=3$, $g_{2,3}(0)=1$, $h_{2,3}(0)=1$ and $f_{2,3}(0)=7$. By the method illustrated in the previous section, we obtain following recursion relations for any non-negative integer $n$.
\beqs
\lefteqn{f_{2,3}(n+1)} \cr\cr
& = & f_{2,3}^6(n) - 3f_{2,3}^3(n)[t_{2,3}(n) + g_{2,3}(n)]^3 - 3f_{2,3}(n)[t_{2,3}(n)+g_{2,3}(n)]^5 \cr\cr 
& & - [t_{2,3}(n)+g_{2,3}(n)]^6 + 6t_{2,3}(n)f_{2,3}(n)[t_{2,3}(n)+g_{2,3}(n)]^4 \cr\cr & & + 6t_{2,3}^2(n)[t_{2,3}(n)+g_{2,3}(n)]^4 - 6t_{2,3}^3(n)[t_{2,3}(n)+g_{2,3}(n)]^3 \ , 
\label{f23eq}
\eeqs
\beqs
\lefteqn{t_{2,3}(n+1)} \cr\cr
& = & 142t_{2,3}^3(n)g_{2,3}^3(n) + 18t_{2,3}^4(n)g_{2,3}(n)h_{2,3}(n) + 45t_{2,3}^3(n)g_{2,3}^2(n)h_{2,3}(n) \cr\cr
& & + 153t_{2,3}^2(n)g_{2,3}^4(n) + 36t_{2,3}^2(n)g_{2,3}^3(n)h_{2,3}(n) + 45t_{2,3}(n)g_{2,3}^5(n) \cr\cr 
& & + 9t_{2,3}(n)g_{2,3}^4(n)h_{2,3}(n)+2g_{2,3}^6(n) \ , 
\label{t23eq}
\eeqs
\beqs
\lefteqn{g_{2,3}(n+1)} \cr\cr
& = & 77t_{2,3}^3(n)g_{2,3}^2(n)h_{2,3}(n) + 171t_{2,3}^2(n)g_{2,3}^4(n) + 2t_{2,3}^4(n)h_{2,3}^2(n) \cr\cr
& & + 22t_{2,3}^3(n)g_{2,3}(n)h_{2,3}^2(n) + 200t_{2,3}^2(n)g_{2,3}^3(n)h_{2,3}(n) + 195t_{2,3}(n)g_{2,3}^5(n) \cr\cr
& & + t_{2,3}^3(n)h_{2,3}^3(n) + 50t_{2,3}^2(n)g_{2,3}^2(n)h_{2,3}^2(n) + 169t_{2,3}(n)g_{2,3}^4(n)h_{2,3}(n) \cr\cr
& & + 56g_{2,3}^6(n) + 3t_{2,3}^2(n)g_{2,3}(n)h_{2,3}^3(n) + 42t_{2,3}(n)g_{2,3}^3(n)h_{2,3}^2(n) \cr\cr
& & + 46g_{2,3}^5(n)h_{2,3} + 3t_{2,3}(n)g_{2,3}^2(n)h_{2,3}^3(n) + 12g_{2,3}^4(n)h_{2,3}^2(n) \cr\cr
& & + g_{2,3}^3(n)h_{2,3}^3(n) \ , 
\label{g23eq}
\eeqs
\beqs
\lefteqn{h_{2,3}(n+1)} \cr\cr
& = & 60t_{2,3}^3(n)g_{2,3}(n)h_{2,3}^2(n) + 564t_{2,3}^2(n)g_{2,3}^3(n)h_{2,3}(n) + 468t_{2,3}(n)g_{2,3}^5(n) \cr\cr & & + 14t_{2,3}^3(n)h_{2,3}^3(n) + 552t_{2,3}^2(n)g_{2,3}^2(n)h_{2,3}^2(n) + 1608t_{2,3}(n)g_{2,3}^4(n)h_{2,3}(n) \cr\cr
& & + 468g_{2,3}^6(n) + 162t_{2,3}^2(n)g_{2,3}(n)h_{2,3}^3(n) + 1404t_{2,3}(n)g_{2,3}^3(n)h_{2,3}^2(n) \cr\cr & & + 1236g_{2,3}^5(n)h_{2,3}(n) + 15t_{2,3}^2(n)h_{2,3}^4(n) + 522t_{2,3}(n)g_{2,3}^2(n)h_{2,3}^3(n) \cr\cr
& & + 1152g_{2,3}^4(n)h_{2,3}^2(n) + 90t_{2,3}(n)g_{2,3}(n)h_{2,3}^4(n)+ 534g_{2,3}^3(n)h_{2,3}^3(n) \cr\cr 
& & + 6t_{2,3}(n)h_{2,3}^5(n) + 135g_{2,3}^2(n)h_{2,3}^4(n) + 18g_{2,3}(n)h_{2,3}^5(n) + h_{2,3}^6(n) \ .
\label{h23eq}
\eeqs
The figures for these configurations are too many to be shown here.
Some values of $f_{2,3}(n)$, $t_{2,3}(n)$, $g_{2,3}(n)$, $h_{2,3}(n)$ are listed in Table \ref{tablesg23}. These numbers grow exponentially, and do not have simple integer factorizations.

\bigskip

\begin{table}[htbp]
\caption{\label{tablesg23} The first few values of $f_{2,3}(n)$, $t_{2,3}(n)$, $g_{2,3}(n)$, $h_{2,3}(n)$.}
\begin{center}
\begin{tabular}{|c||r|r|r|}
\hline\hline 
$n$          & 0 &      1 &                                      2 \\ \hline\hline 
$f_{2,3}(n)$ & 7 & 61,905 & 53,145,523,900,850,102,434,114,604,001 \\ \hline 
$t_{2,3}(n)$ & 3 &  8,372 &    218,891,276,004,139,532,538,695,680 \\ \hline 
$g_{2,3}(n)$ & 1 &  8,020 &  1,242,664,072,161,818,527,545,741,824 \\ \hline 
$h_{2,3}(n)$ & 1 & 29,473 & 49,198,640,408,360,507,318,938,682,849 \\ \hline\hline 
\end{tabular}
\end{center}
\end{table}

\bigskip

By a similar argument as Lemma \ref{lemmasg2b}, the asymptotic growth constant for the number of spanning forests on $SG_{2,3}(n)$ is bounded:
\beq
\frac{5}{7\times 6^m} \ln h_{2,3}(m) < z_{SG_{2,3}} < \frac{5}{7\times 6^m} \ln f_{2,3}(m)  \ ,
\label{zsg23}
\eeq
with $m$ a positive integer. We have the following proposition.

\bigskip

\begin{propo} \label{proposg23} The asymptotic growth constant for the number of spanning forests on the two-dimensional Sierpinski gasket $SG_{2,3}(n)$ in the large $n$ limit is $z_{SG_{2,3}}=1.31235755933...$.

\end{propo}

\bigskip

The convergence of the upper and lower bounds remains quick. By the same method as given in the proof of Proposition \ref{proposg2}, the difference between the upper bound in Eq. (\ref{zsg23}) and the asymptotic growth constant is bounded:
\beqs
\frac{5}{7 \times 6^m} \ln f_{2,3}(m) - z_{SG_{2,3}} & \le & \frac{-1}{7 \times 6^m} \ln \left ( 1-7 \left [\frac{t_{2,3}(m)}{f_{2,3}(m)} + \frac{g_{2,3}(m)}{f_{2,3}(m)} \right ]^3 \right ) \ . \cr & &
\eeqs
More than a hundred significant figures for $z_{SG_{2,3}}$ can be obtained when $m$ is equal to five.

For $SG_{2,4}(n)$, the numbers of edges and vertices are given by 
\beq
e(SG_{2,4}(n)) = 3 \times 10^n \ ,
\label{esg24}
\eeq
\beq
v(SG_{2,4}(n)) = \frac{4 \times 10^n + 5}{3} \ ,
\label{vsg24}
\eeq
where again the three outmost vertices have degree two. There are $(10^n-1)/3$ vertices of $SG_{2,4}(n)$ with degree six, and $(10^n-1)$ vertices with degree four. By Definition \ref{defisg2}, the number of spanning forests is $f_{2,4}(n) = t_{2,4}(n)+3g_{2,4}(n)+h_{2,4}(n)$. The initial values are the same as for $SG_2$: $t_{2,4}(0)=3$, $g_{2,4}(0)=1$, $h_{2,4}(0)=1$ and $f_{2,4}(0)=7$.
We write a computer program to obtain the recursion relations. Using the shorthand notation $tg_{2,4}(n) = t_{2,4}(n) + g_{2,4}(n)$, we have
\beqs
\lefteqn{f_{2,4}(n+1)} \cr\cr
& = & f_{2,4}^{10}(n) - 6f_{2,4}^7(n)tg_{2,4}^3(n) - 9f_{2,4}^5(n)tg_{2,4}^5(n) + 18f_{2,4}^5(n)tg_{2,4}^4(n)t_{2,4}(n) \cr\cr 
& & + 2f_{2,4}^4(n)tg_{2,4}^6(n) + 18f_{2,4}^4(n)tg_{2,4}^4(n)t_{2,4}^2(n) - 18f_{2,4}^4(n)tg_{2,4}^3(n)t_{2,4}^3(n) \cr\cr & & - 6f_{2,4}^3(n)tg_{2,4}^7(n) + 30f_{2,4}^3(n)tg_{2,4}^6(n)t_{2,4}(n) - 30f_{2,4}^3(n)tg_{2,4}^5(n)t_{2,4}^2(n) \cr\cr & & + 3f_{2,4}^2(n)tg_{2,4}^8(n) + 24f_{2,4}^2(n)tg_{2,4}^7(n)t_{2,4}(n) - 36f_{2,4}^2(n)tg_{2,4}^6(n)t_{2,4}^2(n) \cr\cr & & - 54f_{2,4}^2(n)tg_{2,4}^5(n)t_{2,4}^3(n) + 60f_{2,4}^2(n)tg_{2,4}^4(n)t_{2,4}^4(n) - 5f_{2,4}(n)tg_{2,4}^9(n) \cr\cr & & + 42f_{2,4}(n)tg_{2,4}^8(n)t_{2,4}(n) - 42f_{2,4}(n)tg_{2,4}^7(n)t_{2,4}^2(n) \cr\cr & & - 168f_{2,4}(n)tg_{2,4}^6(n)t_{2,4}^3(n) + 330f_{2,4}(n)tg_{2,4}^5(n)t_{2,4}^4(n) \cr\cr & & - 162f_{2,4}(n)tg_{2,4}^4(n)t_{2,4}^5(n) + 8f_{2,4}(n)tg_{2,4}^3(n)t_{2,4}^6(n) + 42tg_{2,4}^8(n)t_{2,4}^2(n) \cr\cr & & - 162tg_{2,4}^7(n)t_{2,4}^3(n) + 102tg_{2,4}^6(n)t_{2,4}^4(n) + 288tg_{2,4}^5(n)t_{2,4}^5(n) \cr\cr & & - 432tg_{2,4}^4(n)t_{2,4}^6(n) + 162tg_{2,4}^3(n)t_{2,4}^7(n) \ .
\label{f24eq}
\eeqs
The other recursion relations for $SG_{2,4}(n)$ are too lengthy to be included here. They are available from the authors on request. Some values of $f_{2,4}(n)$, $t_{2,4}(n)$, $g_{2,4}(n)$, $h_{2,4}(n)$ are listed in Table \ref{tablesg24}. These numbers grow exponentially, and do not have simple integer factorizations.

\bigskip

\begin{table}[htbp]
\caption{\label{tablesg24} The first few values of $f_{2,4}(n)$, $t_{2,4}(n)$, $g_{2,4}(n)$, $h_{2,4}(n)$.}
\begin{center}
\begin{tabular}{|c||r|r|}
\hline\hline 
$n$          & 1 & 2 \\ \hline \hline
$f_{2,4}(n)$ & 75,908,209 & \tiny 6,053,025,303,996,636,848,970,430,785,675,468,144,409,657,412,247,800,423,390,303,465,602,821,564,523,873 \\ \hline 
$t_{2,4}(n)$ &  6,665,475 & \tiny 772,069,425,849,585,011,183,346,692,712,538,703,294,972,628,973,372,161,275,424,155,207,555,217,357 \\ \hline 
$g_{2,4}(n)$ &  8,406,453 & \tiny 17,447,838,129,920,655,302,865,270,986,884,479,355,572,603,291,172,150,410,900,983,156,421,717,259,395 \\ \hline 
$h_{2,4}(n)$ & 44,023,375 & \tiny 5,999,909,720,181,025,298,050,651,626,022,102,167,639,644,629,745,310,599,996,325,091,978,348,857,528,331 \\ \hline\hline 
\end{tabular}
\end{center}
\end{table}

\bigskip

By a similar argument as Lemma \ref{lemmasg2b}, the asymptotic growth constant for the number of spanning forests on $SG_{2,4}(n)$ is bounded:
\beq
\frac{3}{4\times 10^m} \ln h_{2,4}(m) < z_{SG_{2,4}} < \frac{3}{4\times 10^m} \ln f_{2,4}(m)  \ ,
\label{zsg24}
\eeq
with $m$ a positive integer. We have the following proposition.

\bigskip

\begin{propo} \label{proposg24} The asymptotic growth constant for the number of spanning forests on the two-dimensional Sierpinski gasket $SG_{2,4}(n)$ in the large $n$ limit is $z_{SG_{2,4}}=1.36051646575...$.

\end{propo}

\bigskip

The convergence of the upper and lower bounds is again quick. 
By the same method as given in the proof of Proposition \ref{proposg2}, the difference between the upper bound in Eq. (\ref{zsg24}) and the asymptotic growth constant is bounded:
\beqs
\frac{3}{4 \times 10^m} \ln f_{2,4}(m) - z_{SG_{2,4}} & \le & \frac{-1}{12 \times 10^m} \ln \left ( 1-15 \left [\frac{tg_{2,4}(m)}{f_{2,4}(m)} \right ]^3 \right ) \ .
\eeqs
More than a hundred significant figures for $z_{SG_{2,4}}$ can be obtained when $m$ is equal to four.

\section{The number of spanning forests on $SG_d(n)$ with $d=3,4$} 
\label{sectionV}

In this section, we derive the asymptotic growth constant of spanning forests on $SG_d(n)$ with $d=3,4$.
For the three-dimensional Sierpinski gasket $SG_3(n)$, we use the following definitions.

\bigskip

\begin{defi} \label{defisg3} Consider the three-dimensional Sierpinski gasket $SG_3(n)$ at stage $n$. (a) Define $f_3(n) \equiv N_{SF}(SG_3(n))$ as the number of spanning forests. (b) Define $t_3(n)$ as the number of spanning forests such that the four outmost vertices belong to one tree. (c) Define $g_3(n)$ as the number of spanning forests such that one of the outmost vertices belongs to one tree and the other three outmost vertices belong to another tree. (d) Define $h_3(n)$ as the number of spanning forests such that two of the outmost vertices belong to one tree and the other two outmost vertices belong to another tree. (e) Define $p_3(n)$ as the number of spanning forests such that two of the outmost vertices belong to one tree and the other two outmost vertices separately belong to other trees. (f) Define $q_3(n)$ as the number of spanning forests such that each of the outmost vertices belongs to a different tree.
\end{defi}

\bigskip

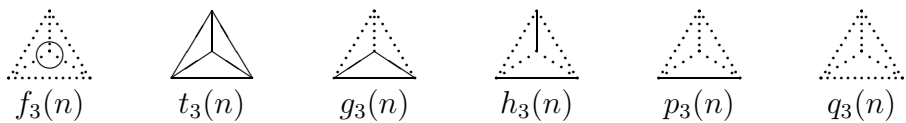
\begin{figure}[htbp]
\unitlength 1.8mm 
\begin{picture}(66,5)
\multiput(0,0)(0.5,0){13}{\circle*{0.2}}
\multiput(0,0)(0.3,0.5){11}{\circle*{0.2}}
\multiput(6,0)(-0.3,0.5){11}{\circle*{0.2}}
\multiput(3,2)(0,0.5){7}{\circle*{0.2}}
\multiput(0,0)(0.6,0.4){6}{\circle*{0.2}}
\multiput(6,0)(-0.6,0.4){6}{\circle*{0.2}}
\put(3,1.7){\circle{2}}
\put(3,-2){\makebox(0,0){$f_3(n)$}}
\put(12,0){\line(1,0){6}}
\put(12,0){\line(3,5){3}}
\put(18,0){\line(-3,5){3}}
\put(12,0){\line(3,2){3}}
\put(18,0){\line(-3,2){3}}
\put(15,2){\line(0,1){3}}
\put(15,-2){\makebox(0,0){$t_3(n)$}}
\put(24,0){\line(1,0){6}}
\multiput(24,0)(0.3,0.5){11}{\circle*{0.2}}
\multiput(30,0)(-0.3,0.5){11}{\circle*{0.2}}
\multiput(27,2)(0,0.5){7}{\circle*{0.2}}
\put(24,0){\line(3,2){3}}
\put(30,0){\line(-3,2){3}}
\put(27,-2){\makebox(0,0){$g_3(n)$}}
\put(36,0){\line(1,0){6}}
\multiput(36,0)(0.3,0.5){11}{\circle*{0.2}}
\multiput(42,0)(-0.3,0.5){11}{\circle*{0.2}}
\put(39,2){\line(0,1){3}}
\multiput(36,0)(0.6,0.4){6}{\circle*{0.2}}
\multiput(42,0)(-0.6,0.4){6}{\circle*{0.2}}
\put(39,-2){\makebox(0,0){$h_3(n)$}}
\put(48,0){\line(1,0){6}}
\multiput(48,0)(0.3,0.5){11}{\circle*{0.2}}
\multiput(54,0)(-0.3,0.5){11}{\circle*{0.2}}
\multiput(51,2)(0,0.5){7}{\circle*{0.2}}
\multiput(48,0)(0.6,0.4){6}{\circle*{0.2}}
\multiput(54,0)(-0.6,0.4){6}{\circle*{0.2}}
\put(51,-2){\makebox(0,0){$p_3(n)$}}
\multiput(60,0)(0.5,0){13}{\circle*{0.2}}
\multiput(60,0)(0.3,0.5){11}{\circle*{0.2}}
\multiput(66,0)(-0.3,0.5){11}{\circle*{0.2}}
\multiput(63,2)(0,0.5){7}{\circle*{0.2}}
\multiput(60,0)(0.6,0.4){6}{\circle*{0.2}}
\multiput(66,0)(-0.6,0.4){6}{\circle*{0.2}}
\put(63,-2){\makebox(0,0){$q_3(n)$}}
\end{picture}

\vspace*{5mm}
\caption{\footnotesize{Illustration for the spanning subgraphs $f_3(n)$, $t_3(n)$, $g_3(n)$, $h_3(n)$, $p_3(n)$ and $q_3(n)$. The two outmost vertices at the ends of a solid line belong to one tree, while the two outmost vertices at the ends of a dot line belong to separated trees.}} 
\label{fghpqfig}
\end{figure}

The quantities $f_3(n)$, $t_3(n)$, $g_3(n)$, $h_3(n)$, $p_3(n)$ and $q_3(n)$ are illustrated in Fig. \ref{fghpqfig}, where only the outmost vertices are shown. There are four equivalent $g_3(n)$, three equivalent $h_3(n)$, and six equivalent $p_3(n)$. By definition,
\beq
f_3(n) = t_3(n)+4g_3(n)+3h_3(n)+6p_3(n)+q_3(n) \ .
\label{fsg3}
\eeq
The initial values at stage zero are $t_3(0)=16$, $g_3(0)=3$, $h_3(0)=1$, $p_3(0)=1$, $q_3(0)=1$ and $f_3(0)=38$. 

\bigskip

\bigskip

The recursion relations are lengthy and given in the appendix. Some values of $f_3(n)$, $t_3(n)$, $g_3(n)$, $h_3(n)$, $p_3(n)$, $q_3(n)$ are listed in Table \ref{tablesg3}. These numbers grow exponentially, and do not have simple integer factorizations.

\bigskip

\begin{table}[htbp]
\caption{\label{tablesg3} The first few values of $f_3(n)$, $t_3(n)$, $g_3(n)$, $h_3(n)$, $p_3(n)$, $q_3(n)$.}
\begin{center}
\begin{tabular}{|c||r|r|r|}
\hline\hline 
$n$      & 0 &      1 &                                      2 \\ \hline\hline 
$f_3(n)$ & 38 & 701,866 & 150,308,440,552,729,541,599,408 \\ \hline 
$t_3(n)$ & 16 & 173,880 &  14,568,001,216,879,127,537,520 \\ \hline 
$g_3(n)$ &  3 &  63,354 &  10,109,099,387,983,187,560,398 \\ \hline 
$h_3(n)$ &  1 &   9,059 &   1,150,970,295,799,746,536,513 \\ \hline
$p_3(n)$ &  1 &  31,357 &   9,282,357,698,529,097,198,747 \\ \hline 
$q_3(n)$ &  1 &  59,251 &  36,156,984,705,343,841,018,275 \\ \hline\hline 
\end{tabular}
\end{center}
\end{table}

\bigskip

By a similar argument as Lemma \ref{lemmasg2b}, the asymptotic growth constant for the number of spanning forests on $SG_3(n)$ is bounded:
\beq
\frac{\ln q_3(m)}{2\times 4^m}  < z_{SG_3} < \frac{\ln f_3(m)}{2\times 4^m}   \ ,
\label{zsg3}
\eeq
with $m$ a positive integer. We have the following proposition.

\bigskip

\begin{propo} \label{proposg3} The asymptotic growth constant for the number of spanning forests on the three-dimensional Sierpinski gasket $SG_3(n)$ in the large $n$ limit is $z_{SG_3}=1.66680628117...$.

\end{propo}

\bigskip

The convergence of the upper and lower bounds is not as quick as for the two dimensional cases. By the same method as given in the proof of Proposition \ref{proposg2}, the difference between the upper bound in Eq. (\ref{zsg3}) and the asymptotic growth constant is bounded:
\beqs
\lefteqn{\frac{1}{2 \times 4^m} \ln f_3(m) - z_{SG_3}} \cr\cr & \le & \frac{-1}{6 \times 4^m} \ln \left ( 1-7 \left [\frac{t_3(m)}{f_3(m)} + \frac{2g_3(m)}{f_3(m)} + \frac{h_3(m)}{f_3(m)} + \frac{p_3(m)}{f_3(m)} \right ]^3 \right ) \ . 
\eeqs
More than a hundred significant figures for $z_{SG_3}$ can be obtained when $m$ is equal to nine.

For the four-dimensional Sierpinski gasket $SG_4(n)$, we use the following definitions.

\bigskip

\begin{defi} \label{defisg4} Consider the four-dimensional Sierpinski gasket $SG_4(n)$ at stage $n$. (a) Define $f_4(n) \equiv N_{SF}(SG_4(n))$ as the number of spanning forests. (b) Define $t_4(n)$ as the number of spanning forests such that the five outmost vertices belong to one tree. (c) Define $g_4(n)$ as the number of spanning forests such that two of the outmost vertices belong to one tree and the other three outmost vertices belong to another tree. (d) Define $h_4(n)$ as the number of spanning forests such that one of the outmost vertices belong to one tree and the other four outmost vertices belong to another tree. (e) Define $p_4(n)$ as the number of spanning forests such that one of the outmost vertices belong to one tree, two of the other outmost vertices belong to another tree and the rest two outmost vertices belong to a third tree. (f) Define $q_4(n)$ as the number of spanning forests such that three of the outmost vertices belong to one tree and the other two outmost vertices separately belong to other trees. (g) Define $r_4(n)$ as the number of spanning forests such that two of the outmost vertices belong to one tree and the other three outmost vertices separately belong to other trees. (h) Define $s_4(n)$ as the number of spanning forests such that each of the outmost vertices belongs to a different tree.
\end{defi}

\bigskip

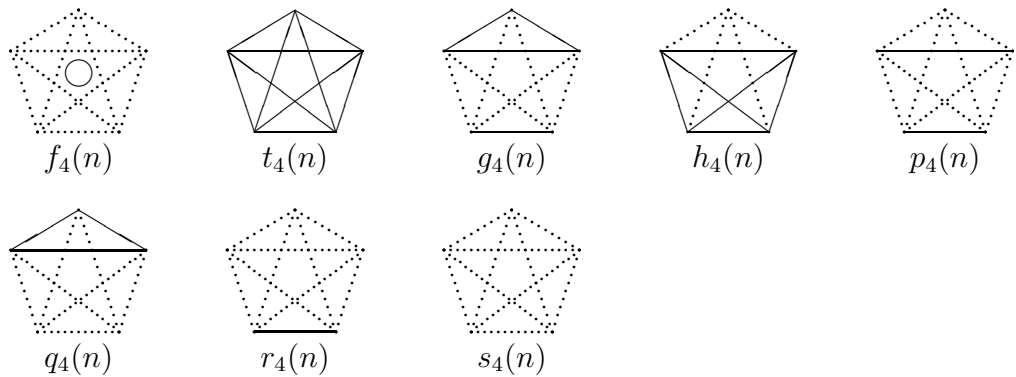
\begin{figure}[htbp]
\unitlength 1.8mm 
\begin{picture}(74,9)
\multiput(2,0)(0.5,0){13}{\circle*{0.2}}
\multiput(2,0)(0.4,0.3){21}{\circle*{0.2}}
\multiput(2,0)(-0.2,0.6){11}{\circle*{0.2}}
\multiput(2,0)(0.2,0.6){16}{\circle*{0.2}}
\multiput(8,0)(-0.2,0.6){16}{\circle*{0.2}}
\multiput(8,0)(0.2,0.6){11}{\circle*{0.2}}
\multiput(8,0)(-0.4,0.3){21}{\circle*{0.2}}
\multiput(0,6)(0.5,0){21}{\circle*{0.2}}
\multiput(5,9)(0.5,-0.3){11}{\circle*{0.2}}
\multiput(5,9)(-0.5,-0.3){11}{\circle*{0.2}}
\put(5,4.4){\circle{2}}
\put(5,-2){\makebox(0,0){$f_4(n)$}}
\put(18,0){\line(1,0){6}}
\put(18,0){\line(4,3){8}}
\put(18,0){\line(-1,3){2}}
\put(18,0){\line(1,3){3}}
\put(24,0){\line(-1,3){3}}
\put(24,0){\line(1,3){2}}
\put(24,0){\line(-4,3){8}}
\put(16,6){\line(1,0){10}}
\put(21,9){\line(5,-3){5}}
\put(21,9){\line(-5,-3){5}}
\put(21,-2){\makebox(0,0){$t_4(n)$}}
\put(34,0){\line(1,0){6}}
\multiput(34,0)(0.4,0.3){21}{\circle*{0.2}}
\multiput(34,0)(-0.2,0.6){11}{\circle*{0.2}}
\multiput(34,0)(0.2,0.6){16}{\circle*{0.2}}
\multiput(40,0)(-0.2,0.6){16}{\circle*{0.2}}
\multiput(40,0)(0.2,0.6){11}{\circle*{0.2}}
\multiput(40,0)(-0.4,0.3){21}{\circle*{0.2}}
\put(32,6){\line(1,0){10}}
\put(37,9){\line(5,-3){5}}
\put(37,9){\line(-5,-3){5}}
\put(37,-2){\makebox(0,0){$g_4(n)$}}
\put(50,0){\line(1,0){6}}
\put(50,0){\line(4,3){8}}
\put(50,0){\line(-1,3){2}}
\multiput(50,0)(0.2,0.6){16}{\circle*{0.2}}
\multiput(56,0)(-0.2,0.6){16}{\circle*{0.2}}
\put(56,0){\line(1,3){2}}
\put(56,0){\line(-4,3){8}}
\put(48,6){\line(1,0){10}}
\multiput(53,9)(0.5,-0.3){11}{\circle*{0.2}}
\multiput(53,9)(-0.5,-0.3){11}{\circle*{0.2}}
\put(53,-2){\makebox(0,0){$h_4(n)$}}
\put(66,0){\line(1,0){6}}
\multiput(66,0)(0.4,0.3){21}{\circle*{0.2}}
\multiput(66,0)(-0.2,0.6){11}{\circle*{0.2}}
\multiput(66,0)(0.2,0.6){16}{\circle*{0.2}}
\multiput(72,0)(-0.2,0.6){16}{\circle*{0.2}}
\multiput(72,0)(0.2,0.6){11}{\circle*{0.2}}
\multiput(72,0)(-0.4,0.3){21}{\circle*{0.2}}
\put(64,6){\line(1,0){10}}
\multiput(69,9)(0.5,-0.3){11}{\circle*{0.2}}
\multiput(69,9)(-0.5,-0.3){11}{\circle*{0.2}}
\put(69,-2){\makebox(0,0){$p_4(n)$}}
\end{picture}

\vspace*{10mm}

\begin{picture}(58,9)
\multiput(2,0)(0.5,0){13}{\circle*{0.2}}
\multiput(2,0)(0.4,0.3){21}{\circle*{0.2}}
\multiput(2,0)(-0.2,0.6){11}{\circle*{0.2}}
\multiput(2,0)(0.2,0.6){16}{\circle*{0.2}}
\multiput(8,0)(-0.2,0.6){16}{\circle*{0.2}}
\multiput(8,0)(0.2,0.6){11}{\circle*{0.2}}
\multiput(8,0)(-0.4,0.3){21}{\circle*{0.2}}
\put(0,6){\line(1,0){10}}
\put(5,9){\line(5,-3){5}}
\put(5,9){\line(-5,-3){5}}
\put(5,-2){\makebox(0,0){$q_4(n)$}}
\put(18,0){\line(1,0){6}}
\multiput(18,0)(0.4,0.3){21}{\circle*{0.2}}
\multiput(18,0)(-0.2,0.6){11}{\circle*{0.2}}
\multiput(18,0)(0.2,0.6){16}{\circle*{0.2}}
\multiput(24,0)(-0.2,0.6){16}{\circle*{0.2}}
\multiput(24,0)(0.2,0.6){11}{\circle*{0.2}}
\multiput(24,0)(-0.4,0.3){21}{\circle*{0.2}}
\multiput(16,6)(0.5,0){21}{\circle*{0.2}}
\multiput(21,9)(0.5,-0.3){11}{\circle*{0.2}}
\multiput(21,9)(-0.5,-0.3){11}{\circle*{0.2}}
\put(21,-2){\makebox(0,0){$r_4(n)$}}
\multiput(34,0)(0.5,0){13}{\circle*{0.2}}
\multiput(34,0)(0.4,0.3){21}{\circle*{0.2}}
\multiput(34,0)(-0.2,0.6){11}{\circle*{0.2}}
\multiput(34,0)(0.2,0.6){16}{\circle*{0.2}}
\multiput(40,0)(-0.2,0.6){16}{\circle*{0.2}}
\multiput(40,0)(0.2,0.6){11}{\circle*{0.2}}
\multiput(40,0)(-0.4,0.3){21}{\circle*{0.2}}
\multiput(32,6)(0.5,0){21}{\circle*{0.2}}
\multiput(37,9)(0.5,-0.3){11}{\circle*{0.2}}
\multiput(37,9)(-0.5,-0.3){11}{\circle*{0.2}}
\put(37,-2){\makebox(0,0){$s_4(n)$}}
\end{picture}

\vspace*{5mm}
\caption{\footnotesize{Illustration for the spanning subgraphs $f_4(n)$, $t_4(n)$, $g_4(n)$, $h_4(n)$, $p_4(n)$, $q_4(n)$, $r_4(n)$ and $s_4(n)$. The two outmost vertices at the ends of a solid line belong to one tree, while the two outmost vertices at the ends of a dot line belong to separated trees.}} 
\label{fghpqrsfig}
\end{figure}

The quantities $f_4(n)$, $t_4(n)$, $g_4(n)$, $h_4(n)$, $p_4(n)$, $q_4(n)$, $r_4(n)$ and $s_4(n)$ are illustrated in Fig. \ref{fghpqrsfig}, where only the outmost vertices are shown. There are ten equivalent $g_4(n)$, five equivalent $h_4(n)$, fifteen equivalent $p_4(n)$, ten equivalent $q_4(n)$ and ten equivalent $r_4(n)$. By definition,
\beqs
f_4(n) & = & t_4(n)+10g_4(n)+5h_4(n)+15p_4(n)+10q_4(n)+10r_4(n)+s_4(n) \ . \cr & &
\label{fsg4}
\eeqs
The initial values at stage zero are $t_4(0)=125$, $g_4(0)=3$, $h_4(0)=16$, $p_4(0)=1$, $q_4(0)=3$, $r_4(0)=1$, $s_4(0)=1$ and $f_4(0)=291$.

\bigskip

\bigskip

We write a computer program to obtain the recursion relations. Using the shorthand notations $tr_4(n) = t_4(n) + 4g_4(n) + 3h_4(n) + 3p_4(n) + 3q_4(n) + r_4(n)$, $tq_4(n) = t_4(n) + g_4(n) + 2h_4(n) + q_4(n)$, $tp_4(n) = t_4(n) + 2g_4(n) + h_4(n) + p_4(n)$ and $th_4(n) = t_4(n) + h_4(n)$, we have
\beqs
\lefteqn{f_4(n+1)} \cr\cr
& = & f_4^5(n) - 10f_4^2(n)tr_4^3(n) - 15f_4(n)tr_4^4(n) - 30f_4(n)tq_4^4(n) - 12tr_4^5(n) \cr\cr 
& & + 60f_4(n)tr_4^2(n)tq_4^2(n) - 15tr_4^4(n)tp_4(n) + 30tr_4^4(n)th_4(n) \cr\cr & & - 30tr_4^3(n)tp_4^2(n) + 120tr_4^3(n)tp_4(n)th_4(n) + 140tr_4^3(n)tq_4^2(n) \cr\cr & & - 120tr_4^3(n)th_4^2(n) + 240tr_4^2(n)tq_4^2(n)tp_4(n) - 480tr_4^2(n)tq_4^2(n)th_4(n) \cr\cr & & + 300tr_4(n)tq_4^2(n)tp_4^2(n) - 1200tr_4(n)tq_4^2(n)tp_4(n)th_4(n) \cr\cr & & - 180tr_4(n)tq_4^4(n) + 1200tr_4(n)tq_4^2(n)th_4^2(n) - 51tp_4^5(n) \cr\cr & & + 510tp_4^4(n)th_4(n) + 260tq_4^2(n)tp_4^3(n) - 2040tp_4^3(n)th_4^2(n) \cr\cr & & - 1560tq_4^2(n)tp_4^2(n)th_4(n) + 4080tp_4^2(n)th_4^3(n) - 210tq_4^4(n)tp_4(n) \cr\cr & & + 3120tq_4^2(n)tp_4(n)th_4^2(n) - 4080tp_4(n)th_4^4(n) + 420tq_4^4(n)th_4(n) \cr\cr & & - 2080tq_4^2(n)th_4^3(n) + 1632th_4^5(n)  \ .
\label{f4eq}
\eeqs
The other recursion relations for $SG_4(n)$ are too lengthy to be included here. They are available from the authors on request. Some values of $f_4(n)$, $t_4(n)$, $g_4(n)$, $h_4(n)$, $p_4(n)$, $q_4(n)$, $r_4(n)$, $s_4(n)$ are listed in Table \ref{tablesg4}. These numbers grow exponentially, and do not have simple integer factorizations.

\bigskip

\begin{table}[htbp]
\caption{\label{tablesg4} The first few values of $f_4(n)$, $t_4(n)$, $g_4(n)$, $h_4(n)$, $p_4(n)$, $q_4(n)$, $r_4(n)$, $s_4(n)$.}
\begin{center}
\begin{tabular}{|c||r|r|}
\hline\hline 
$n$          & 1 & 2 \\ \hline \hline
$f_4(n)$ & 85,824,132,029 & \footnotesize 7,035,17,527,028,105,500,700,677,412,563,863,619,648,991,055,157,831,483 \\ \hline 
$t_4(n)$ &  3,412,986,435 & \footnotesize 96,263,552,482,319,683,899,326,687,304,651,572,426,360,843,549,870,965 \\ \hline 
$g_4(n)$ &    392,122,089 & \footnotesize 2,066,883,222,491,708,347,294,489,449,954,683,350,540,164,424,914,435 \\ \hline 
$h_4(n)$ &  5,923,774,096 & \footnotesize 40,841,537,587,690,687,322,887,835,686,137,425,636,710,177,922,212,520 \\ \hline
$p_4(n)$ &    224,652,411 & \footnotesize 1,952,486,255,633,069,494,764,677,365,066,319,434,639,193,908,980,317 \\ \hline 
$q_4(n)$ &  1,740,690,487 & \footnotesize 19,621,800,909,697,266,778,177,200,667,594,598,639,201,513,851,821,683 \\ \hline 
$r_4(n)$ &    693,438,141 & \footnotesize 12,210,477,454,458,190,580,945,663,798,559,596,025,810,422,699,074,029 \\ \hline 
$s_4(n)$ &  1,159,981,779 & \footnotesize 34,767,376,906,364,680,701,267,847,191,441,347,363,970,403,604,091,693 \\ \hline 
\hline\hline 
\end{tabular}
\end{center}
\end{table}

\bigskip

By a similar argument as Lemma \ref{lemmasg2b}, the asymptotic growth constant for the number of spanning forests on $SG_4(n)$ is bounded:
\beq
\frac{2}{5^{m+1}} \ln s_4(m) < z_{SG_4} < \frac{2}{5^{m+1}} \ln f_4(m)  \ ,
\label{zsg4}
\eeq
with $m$ a positive integer. We have the following proposition.

\bigskip

\begin{propo} \label{proposg4} The asymptotic growth constant for the number of spanning forests on the two-dimensional Sierpinski gasket $SG_4(n)$ in the large $n$ limit is $z_{SG_4}=1.98101707560...$.

\end{propo}

\bigskip

The convergence of the upper and lower bounds is even slower compared with that for $SG_3$. By the same method as given in the proof of Proposition \ref{proposg2}, the difference between the upper bound in Eq. (\ref{zsg4}) and the asymptotic growth constant is bounded:
\beqs
\frac{2}{5^{m+1}} \ln f_4(m) - z_{SG_4} & \le & \frac{-1}{2 \times 5^{m+1}} \ln \left ( 1-67 \left [\frac{tr_4(m)}{f_4(m)} \right ]^3 \right ) \ . 
\eeqs

We only have fourteen significant figures for $z_{SG_4}$ with $m$ calculated up to six.

\section{Bounds of the asymptotic growth constants}

As the spanning tree is a special case of spanning forest where there is only one component, it is clear that the number of spanning trees $N_{ST}(G)$ is always less than $N_{SF}(G)$. Define
\beq
\underline z_G = \lim_{v(G) \to \infty} \frac{\ln N_{ST}(G)}{v(G)} \ ,
\label{zdefn}
\eeq
then $\underline z_G < z_G$. We have obtained such asymptotic growth constants for the number of spanning trees on the Sierpinski gasket $SG_d$ for general $d$ and $SG_{2,b}$ with $b=3,4$ in Ref. \cite{sts}. They serve as the lower bounds for our current consideration for the spanning forests.

By Eq. (\ref{v}) and a similar argument as Lemma \ref{lemmasg2b}, we have the upper bound of the asymptotic growth constant for the number of spanning forests on $SG_d(n)$:
\beq
z_{SG_d} < \frac{2}{(d+1)^{m+1}} \ln N_{SF}(SG_d(m))  \ ,
\label{zsgd}
\eeq
with $m$ a positive integer. Although the number $N_{SF}(SG_d(m))$ for general $m$ is difficult to obtain, it is known for $m=0$. We first recall that $SG_d(0)$ at stage zero is a complete graph with $(d+1)$ vertices, each of which is adjacent to all of the other vertices. The number of spanning forests on the complete graph is given by sequence A001858 in Ref. \cite{sl}. The first few values of $N_{SF}(SG_d(0))$ are 7, 38, 291, 2932 for $d$ from 2 to 5 \cite{callan03}. Define
\beq
\bar z_{SG_d} = \frac{2}{d+1} \ln N_{SF}(SG_d(0))  \ ,
\label{zsgdn}
\eeq
then $z_{SG_d} < \bar z_{SG_d}$. We list the first few values of $\underline z_{SG_d}$, $z_{SG_d}$, $\bar z_{SG_d}$ and their ratios in Table \ref{zsgdtable}. Notice that the upper bound is closer to the exact value when $d$ is small, while the lower bound is closer to the exact value when $d$ is large.

\bigskip

\begin{table}
\caption{\label{zsgdtable} Numerical values of $\underline z_{SG_d}$, $z_{SG_d}$, $\bar z_{SG_d}$ and their ratios. The last digits given are rounded off.}
\begin{center}
\begin{tabular}{|c|c|c|c|c|c|c|}
\hline\hline 
$d$ &  $D$  & $\underline z_{SG_d}$ & $z_{SG_d}$ & $\bar z_{SG_d}$ & $\underline z_{SG_d}/z_{SG_d}$ & $z_{SG_d}/\bar z_{SG_d}$ \\ \hline\hline 
2   & 1.585 & 1.048594857 & 1.247337199 & 1.297273433 & 0.8406667076 & 0.9615067787 \\ \hline
3   & 2     & 1.569396409 & 1.666806281 & 1.818793080 & 0.9415589724 & 0.9164353546 \\ \hline
4   & 2.322 & 1.914853265 & 1.981017076 & 2.269329307 & 0.9666010902 & 0.8729526691 \\ \hline
5   & 2.585 & 2.172764568 & -           & 2.661146688 & -            & - \\ \hline\hline 
\end{tabular}
\end{center}
\end{table}

\bigskip

For the generalized Sierpinski gasket $SG_{2,b}(n)$ with dimension equal to two, the number of vertices can be calculated to be
\beq
v(SG_{2,b}(n)) = \frac{b+4}{b+2} \left[ \frac{b(b+1)}{2} \right ]^n + \frac{2(b+1)}{b+2} \ .
\eeq
The upper bound of the asymptotic growth constant for the number of spanning forests on $SG_{2,b}(n)$ is given by
\beq
z_{SG_{2,b}} < \left ( \frac{b+2}{b+4} \right ) \frac{\ln N_{SF}(SG_{2,b}(m))}{\left[b(b+1)/2 \right]^m}   \ ,
\label{zsg2b}
\eeq
with $m$ a positive integer. Although the number $N_{SF}(SG_{2,b}(m))$ for general $m$ is difficult to obtain, it is always equal to seven for stage zero since $SG_{2,b}(0)$ is the equilateral triangle. Define 
\beq
\bar z_{SG_{2,b}} = \frac{b+2}{b+4} \ln 7  \ ,
\label{zsg2bn}
\eeq
then $z_{SG_{2,b}} < \bar z_{SG_{2,b}}$. We list the first few values of $\underline z_{SG_{2,b}}$, $z_{SG_{2,b}}$, $\bar z_{SG_{2,b}}$ and their ratios in Table \ref{zsg2btable}. Notice that the upper bound is closer to the exact value when $b$ is small, while the lower bound is closer to the exact value when $b$ is large.

\bigskip

\begin{table}
\caption{\label{zsg2btable} Numerical values of $\underline z_{SG_{2,b}}$, $z_{SG_{2,b}}$, $\bar z_{SG_{2,b}}$ and their ratios. The last digits given are rounded off.}
\begin{center}
\begin{tabular}{|c|c|c|c|c|c|c|}
\hline\hline 
$b$      &  $D$  & $\underline z_{SG_{2,b}}$ & $z_{SG_{2,b}}$ & $\bar z_{SG_{2,b}}$ & $\underline z_{SG_{2,b}}/z_{SG_{2,b}}$ & $z_{SG_{2,b}}/\bar z_{SG_{2,b}}$  \\ \hline\hline 
3        & 1.631 & 1.133231895 & 1.312357559 & 1.389935821 & 0.8635084908 & 0.9441857241 \\ \hline
4        & 1.661 & 1.194401491 & 1.360516466 & 1.459432612 & 0.8779030028 & 0.9322228754 \\ \hline
$\infty$ & 2     & -           & -           & 1.945910149 & - & - \\
\hline\hline 
\end{tabular}
\end{center}
\end{table}

\bigskip


\appendix

\section{Recursion relations for $SG_3(n)$}

We give the recursion relations for the three-dimensional Sierpinski gasket $SG_3(n)$ here. For any non-negative integer $n$, we have
\beqs
\lefteqn{f_3(n+1)} \cr\cr
& = & f_3^4(n) - 4f_3(n)[t_3(n)+2g_3(n)+h_3(n)+p_3(n)]^3 
\cr\cr
& & - 3[t_3(n)+2g_3(n)+h_3(n)+p_3(n)]^4 \cr\cr
& & + 12[t_3(n)+2g_3(n)+h_3(n)+p_3(n)]^2[t_3(n)+g_3(n)]^2 \cr\cr
& & - 6[t_3(n)+g_3(n)]^4 \ , 
\label{f3eq}
\eeqs
\beqs
\lefteqn{t_3(n+1)} \cr\cr
& = & 72t_3^2(n)p_3(n)[g_3(n)+h_3(n)] + 56t_3(n)[g_3(n)+h_3(n)]^3 + 24t_3^2(n)p_3^2(n) \cr\cr
& & + 12t_3(n)p_3(n)[11g_3^2(n)+12g_3(n)h_3(n)+h_3^2(n)] \cr\cr 
& & + 12g_3^2(n)[3g_3^2(n)+8g_3(n)h_3(n)+6h_3^2(n)] \cr\cr
& & + 12t_3(n)p_3^2(n)[4g_3(n)+h_3(n)] + 48g_3^2(n)p_3(n)[g_3(n)+h_3(n)] \cr\cr 
& & + 4t_3(n)p_3^3(n) + 12g_3^2(n)p_3^2(n) \ , 
\label{t3eq}
\eeqs
\beqs
\lefteqn{g_3(n+1)} \cr\cr
& = & 6t_3^2(n)q_3(n)[g_3(n)+h_3(n)] + 24t_3^2(n)p_3^2(n) \cr\cr
& & + 12t_3(n)p_3(n)[9g_3^2(n)+16g_3(n)h_3(n)+7h_3^2(n)] \cr\cr
& & + 4g_3(n)[5g_3^3(n)+18g_3^2(n)h_3(n)+24g_3(n)h_3^2(n)+14h_3^3(n)] \cr\cr
& & + 6t_3^2(n)p_3(n)q_3(n) + 6t_3(n)p_3^2(n)[21g_3(n)+11h_3(n)] \cr\cr 
& & + 3t_3(n)q_3(n)[5g_3^2(n)+6g_3(n)h_3(n)+h_3^2(n)] \cr\cr
& & + 24g_3(n)p_3(n)[5g_3^2(n)+9g_3(n)h_3(n)+4h_3^2(n)] \cr\cr
& & + 6t_3(n)p_3(n)q_3(n)[3g_3(n)+h_3(n)] + 21t_3(n)p_3^3(n) \cr\cr 
& & + 6g_3(n)p_3^2(n)[19g_3(n)+13h_3(n)] \cr\cr
& & + 3g_3(n)q_3(n)[3g_3^2(n)+4g_3(n)h_3(n)+h_3^2(n)] + 3t_3(n)p_3^2(n)q_3(n) \cr\cr
& & + 6g_3(n)p_3(n)q_3(n)[2g_3(n)+h_3(n)] + 25g_3(n)p_3^3(n) \cr\cr
& & + 3g_3(n)p_3^2(n)q_3(n) \ ,
\label{g3eq}
\eeqs
\beqs
\lefteqn{h_3(n+1)} \cr\cr
& = & 2t_3^2(n)p_3^2(n) + 12t_3(n)p_3(n)[g_3^2(n)+4g_3(n)h_3(n)+3h_3^2(n)] \cr\cr
& & + 2[g_3^4(n)+8g_3^3(n)h_3(n)+18g_3^2(n)h_3^2(n)+16g_3(n)h_3^3(n)+11h_3^4(n)] \cr\cr
& & + 8t_3(n)p_3^2(n)[2g_3(n)+3h_3(n)] \cr\cr 
& & + 8p_3(n)[2g_3^3(n)+9g_3^2(n)h_3(n)+9g_3(n)h_3^2(n)+2h_3^3(n)] \cr\cr
& & + 4t_3(n)p_3^3(n) + 2p_3^2(n)[10g_3^2(n)+24g_3(n)h_3(n)+9h_3^2(n)] \cr\cr
& & + 8p_3^3(n)[g_3(n)+h_3(n)] + p_3^4(n)  \ , 
\label{h3eq}
\eeqs
\beqs
\lefteqn{p_3(n+1)} \cr\cr
& = & 6t_3^2(n)p_3(n)q_3(n) + 120t_3(n)p_3^2(n)[g_3(n)+h_3(n)] \cr\cr
& & + 14t_3(n)q_3(n)[g_3(n)+h_3(n)]^2 + 88p_3(n)[g_3(n)+h_3(n)]^3 \cr\cr
& & + 4t_3(n)p_3(n)q_3(n)[13g_3(n)+10h_3(n)] + 78t_3(n)p_3^3(n) \cr\cr
& & + 6p_3^2(n)[49g_3^2(n)+78g_3(n)h_3(n)+29h_3^2(n)] \cr\cr
& & + 2q_3(n)[11g_3^3(n)+26g_3^2(n)h_3(n)+19g_3(n)h_3^2(n)+4h_3^3(n)] \cr\cr
& & + t_3^2(n)q_3^2(n) + 2t_3(n)q_3^2(n)[2g_3(n)+h_3(n)] + 26t_3(n)p_3^2(n)q_3(n) \cr\cr
& & + 2p_3(n)q_3(n)[38g_3^2(n)+50g_3(n)h_3(n)+15h_3^2(n)] \cr\cr
& & + 2p_3^3(n)[115g_3(n)+76h_3(n)] + 2t_3(n)p_3(n)q_3^2(n) \cr\cr
& & + q_3^2(n)[4g_3^2(n)+4g_3(n)h_3(n)+h_3^2(n)] \cr\cr
& & + 2p_3^2(n)q_3(n)[31g_3(n)+18h_3(n)] + 49p_3^4(n) \cr\cr
& & + 2p_3(n)q_3^2(n)[2g_3(n)+h_3(n)] + 14p_3^3(n)q_3(n) + p_3^2(n)q_3^2(n) \ , 
\label{p3eq}
\eeqs
\beqs
\lefteqn{q_3(n+1)} \cr\cr
& = & 144t_3(n)p_3(n)q_3(n)[g_3(n)+h_3(n)] + 208t_3(n)p_3^3(n) \cr\cr
& & + 720p_3^2(n)[g_3(n)+h_3(n)]^2 + 56q_3(n)[g_3(n)+h_3(n)]^3 \cr\cr
& & + 24t_3(n)q_3^2(n)[g_3(n)+h_3(n)] + 252t_3(n)p_3^2(n)q_3(n) \cr\cr
& & + 24p_3(n)q_3(n)[25g_3^2(n)+44g_3(n)h_3(n)+19h_3^2(n)] \cr\cr
& & + 104p_3^3(n)[17g_3(n)+15h_3(n)] + 60t_3(n)p_3(n)q_3^2(n) \cr\cr
& & + 24q_3^2(n)[3g_3^2(n)+5g_3(n)h_3(n)+2h_3^2(n)] \cr\cr
& & + 12p_3^2(n)q_3(n)[110g_3(n)+89h_3(n)] + 972p_3^4(n) + 4t_3(n)q_3^3(n) \cr\cr
& & + 12p_3(n)q_3^2(n)[22g_3(n)+17h_3(n)] + 776p_3^3(n)q_3(n) \cr\cr
& & + 4q_3^3(n)[4g_3(n)+3h_3(n)] + 210p_3^2(n)q_3^2(n) + 24p_3(n)q_3^3(n) + q_3^4(n) \ . \cr & &
\label{q3eq}
\eeqs

\bigskip

\subsection*{Acknowledgment}

The research of S.C.C. was partially supported by the NSC grant NSC-95-2112-M-006-004. The research of L.C.C was partially supported by the NSC grant NSC-95-2115-M-030-002.

\end{document}